\documentclass[12pt]{iopartmod}

\usepackage{eucal}

\def \Diff #1#2 {\frac{\partial #1}{\partial #2} }
\def \Viff #1#2 {\frac{\delta #1}{\delta #2} }
\def \diff #1#2 { #1_{, #2}}
\def \pdiff #1 {\partial_#1}

\def \tDiff #1#2 {\frac{\tD #1}{\tD #2} }
\def \tdiff #1#2 {\frac{\td #1}{\td #2} }

\def\td{ d}
\def\tD{ D}
\def\Lapl {{\triangle}}                       

\def \Chr #1#2#3 {\Gamma^#1_{\: #2#3} }

\def\tnh{\mathop{\rm th}\nolimits}

\def\csh{\mathop{\rm ch}\nolimits}

\def\sh{\mathop{\rm sh}\nolimits}
\def\ch{\mathop{\rm ch}\nolimits}

\def\arctanh{\mathop{\rm arctanh}\nolimits}
\def\sign{\mathop{\rm sign}\nolimits}

\def\cotan{\mathop{\rm cotan}\nolimits}

\def\Gj{\mathop{\rm Gj}\nolimits}

\def \varchi{ {\raise2.5pt\hbox{$\chi$}}             }
\def \svarchi{{\raise1.5pt\hbox{$\scriptstyle\chi$}} }

\def \F{\mathop{\strut\rm F}\nolimits}  
\def \E{\mathop{\strut  \rm E}\nolimits}  

\def \eqhat{\mathop{ \,\hat =\,}}

\def \cma {\,,}
\def \cmb {\,.}
\def \cmma {\ \,,}
\def \cmmb {\ \,.}

\def \scri {\mathcal{J}}

\def \dz #1#2 {(z_{#1}{-}z_{#2})}
\def \dw #1#2 {w_{#1}{-}w_{#2}}

\def \<{\hspace{-2pt}}  

\def \ds{\displaystyle}
\def \scs{\scriptstyle}
\def \txts{\textstyle}

\usepackage[dvips]{graphicx}
\usepackage[center]{subfigure}

\usepackage{bm}
\usepackage{amsmath}
\usepackage{amsthm}
\usepackage{amssymb}

\def \Gj {G\!j}

\def \S {{\small \text{\!\! \it S\hskip0.1em}}}

\font\smalltesti=cmmi8 at 8.0pt
\font\smallseveni=cmmi7 at 5.83pt
\def\K{{\fam=15 K}}

\begin{document}
\title{Asymptotic properties of the C-Metric}
\author{P Sl\'adek$^1$ and J D Finley, III$^2$}
\address{$^1$ Institute of Theoretical Physics, Charles University in Prague,
V Hole\v{s}ovi\v{c}k\'ach 2, 18000 Prague 8, Czech Republic}
\address{$^2$ Department of Physics and Astronomy, University of New Mexico,
Albuquerque, NM 87131 U.S.A.}
\eads{\mailto{sladek@utf.mff.cuni.cz} and \mailto{finley@phys.unm.edu}}

\begin{abstract}
The aim of this article is to analyze the asymptotic properties of the C-metric, using
 a general method specified in work of Tafel and coworkers,
 \cite{tafel_1}, \cite{tafel_2}, \cite{tafel_3}. By finding
 an appropriate conformal factor $\Omega$, it allows the investigation of the asymptotic
 properties of a given asymptotically flat spacetime. The news function and Bondi mass
 aspect are computed, their general properties are analyzed, as well as the small mass, small
 acceleration, small and large Bondi time limits.
\end{abstract}

\maketitle

\section{Introduction}
\label{introduction}

The C-metric is commonly regarded as a spacetime describing two black
holes accelerating in opposite directions, under the action of forces represented
by conical singularities; see
e.g. \cite{cornish}, \cite{kinnersley}, \cite{self_1},
 \cite{krt}. It admits boost and rotational symmetry and belongs thus to the large class of boost-rotation symmetric
 spacetimes \cite{bicak_2}.
While not strictly\footnote{Two generators of $\scri$ are missing, but the spacetime is still
locally asymptotically flat, see \cite{ashtekar_1}; in fact for some parameter values the $\scri^+$ of the C-metric
even allows some global spherical sections, see also \cite{bicak_2}.}  asymptotically flat,
the Bondi coordinates still exist and the news function and mass aspect can be computed.
One can start from the news function and asymptotic properties of a general boost-rotation symmetric spacetime
(see equation (26) in \cite{bicak_1}, (17) in \cite{BiBo}, or (99) and (121) in \cite{BiPra}).
However, here we shall adopt another method, and will give the final expression in the explicit form specific to the
C-metric. The method created by Tafel and coworkers \cite{tafel_1}, \cite{tafel_2}, \cite{tafel_3} is especially
useful for such a computation 
 since, unlike the traditional
approach of finding an asymptotic form
of a coordinate transformation to Bondi coordinates by comparing expansions of the metric
tensor components, it provides a technique to find a specifically ``calibrated'' form of Penrose's
conformal factor, from which those quantities can be directly computed.  These authors follow
(a weakened version of) Penrose's definition of asymptotic flatness\cite{Penrose1}, \cite{Penrose2} to obtain a
particular embedding of a given manifold into an unphysical, conformal compactification.  By
making a set of constraints on the conformal factor, it can be made quite suitable for asymptotic
analysis, so that it and its various derivatives can be used to determine the desired news
function, Bondi mass, and other asymptotic properties.  Our goal here is to follow their format,
determining a conformal factor that satisfies their ``calibration conditions,"
as applied to the C-metric, in appropriate coordinates, to obtain the desired asymptotic properties.

The definition used in \cite{tafel_1}, \cite{tafel_2}, \cite{tafel_3} says that a spacetime,
$\widetilde{\cal M}$, with metric $\widetilde{g}$,
is asymptotically flat at $\scri^+$
(future null infinity) if and only if the following assumptions are satisfied.
\begin{enumerate}\label{OmegaConditions}
\item[(a)] the physical spacetime $\widetilde{\cal M}$ is a submanifold of an unphysical spacetime
 $\cal M$.  The metric, $g$, of $\cal M$ is conformally equivalent on $\widetilde{\cal M}$ to the
metric $\widetilde g$ of $\widetilde{\cal M}$, i.e., we have a conformal transformation induced
by the function $\Omega$, which is required to always be positive on $\widetilde{\cal M}$:
\begin{equation} g = \Omega^2{\widetilde g}\;,
\end{equation}
\item[(b)] A boundary of $\widetilde{\cal M}$ in $\cal M$ contains a three-dimensional null surface
$\scri^+$ such that $\Omega$ vanishes on that boundary although $d\Omega$ must not, as well as a
third condition on the derivative. Using the symbol
$\hat=$ to denote that something is being evaluated on that boundary, we may write these conditions
explicitly:
\begin{equation} \Omega\, \hat=\, 0\;,\quad d\Omega\, \hat{\neq}\, 0\;,
\hbox{~and~~}g^{\mu\nu}\Omega_{,\mu}\Omega_{,\nu}\, \hat=\, 0\;.
\end{equation}
\item[(c)] The boundary $\scri^+$ is diffeomorphic to the product $R\times S_2$.  Thought of as a
trivial bundle over $S_2$ the fibres are generated by the future-directed vector field
\begin{equation}
v = g^{\mu\nu}\Omega_{,\nu}\partial_\mu\;.
\end{equation}
\item[(d)] The pullback of $g$ under the natural embedding $\phi: \scri^+\rightarrow{\cal M}$ is
the natural metric on the sphere:
\begin{equation}
\phi^*g = g_s\;.
\end{equation}
\item[(e)] The Ricci tensor of $\tilde g$ satisfies (in coordinates of ${\cal M}$) the boundary condition :
\begin{equation}
\widetilde R_{\mu\nu}\,  \hat=\, 2q \, \Omega_{,\mu} \Omega_{,\nu} \cma
\end{equation}
where $q$ is a function.
\end{enumerate}
  The method is then to follow a sequence of (allowed) transformations of the natural choice for
the conformal factor $\Omega$, which arranges in turn for the satisfaction of these requirements,
as we will describe in detail below for our metric of interest, a particular choice of the range
of variables for the C-metric that can be described as two black holes accelerating oppositely.  This
will also involve various changes of coordinates, heading toward coordinates that have the
Bondi-Sachs form.\cite{Bondi}\cite{Sachs}
(We also note that the news function for the C-metric has been computed in some earlier works,
see e.g. \cite{farhoosh} and general results in \cite{BiBo}, \cite{BiRo}, \cite{BiPra}, using quite a different method, which
can be used as an independent check of that portion of the results here.)

In the following text, we always use the metric signature $(-+++)$, unlike in \cite{tafel_1},
 \cite{tafel_2}, \cite{tafel_3};
this will often lead to associated sign changes in the equations used and referenced herein.
Raising and lowering of the indices,
and the covariant derivative, denoted by the symbols $_{|\mu}$, are all
understood to be with respect to the unphysical metric $g=\Omega^2 \tilde g$, unless specified otherwise.
The symbol $\hat=$ is used in the same sense as above, i.e. to denote that the expression is being
evaluated on the $\scri^+$ boundary.

Once the final conformal factor $\Omega\, {\equiv}\, \Omega_F$ is obtained, it would allow us to compute the
Bondi mass aspect $M$ and the news function $c_{,u}$, important quantities
characterizing the asymptotic properties of the spacetime, namely the total mass $m(u)$
and its change :
\begin{align}\notag\label{bondiQ}
m(u)&=\frac{1}{4\pi}\int_{S_2}  M(u,\Theta,\Phi) \, dS  = \frac{1}{4\pi}\int_{S_2} \hat M(u,\Theta,\Phi) \, dS \cma  \\
& m_{,u}(u) = - \frac{1}{4\pi}\int_{S_2} \! c_{,u}^2(u,\Theta,\Phi) \, dS \cmma
\end{align}
where $\hat M$ is the so called modified mass aspect and the integral is to be taken over a constant $u$ slice of $\scri^+$, i.e. a unit two-sphere.
The modified mass aspect differs from $M$ by a four-divergence constructed from the news tensor (see
\cite{tafel_2}(72)), thus allowing us to use it in the integral instead of $M$.
The advantage of $\hat M$ lies in its simpler behaviour under the general BMS group, see \cite{tafel_2}(36) and
below. Also it is preferred in the construction of the asymptotic four-momentum, which has
then correct transformation properties under the full BMS group, and not just under the Lorentz
subgroup (see equations (7),(8) in \cite{tafel_2} for more details).

The BMS (Bondi-Metzner-Sachs) group mentioned above is the group of the asymptotic symmetries for any
asymptotically flat spacetime, also inducing a corresponding mapping of the $\scri^+$ into
itself. Notably, it is a much larger group than the Poincar\'e group, since it contains
an infinite dimensional abelian normal subgroup, the supertranslations. The factor group obtained by quotienting
it out is then isomorphic to the Lorentz group, see \cite{wald}.
The action of any supertranslation on $\scri^+$ manifests itself as an angle dependent translation of the Bondi time, $u
\rightarrow u + \alpha(\Theta,\Phi)$. This will become useful
for construction of the Schwarzchild limit in section \ref{SchwLim}.

The news function $c_{,u}$ and the Bondi mass aspect $M$ also appear in the asymptotic
metric expansion in Bondi coordinates.
 For the simplest case of axial symmetry, without an electromagnetic field and without
rotation, the Bondi metric reads (see e.g. \cite{Bondi}):
\begin{align}\label{BondiMetric}
g_B= \Big( {-}\frac{V}{r} e^{2\beta} + & U^2r^2 e^{2\gamma} \Big) du^2 - 2e^{2\beta} dudr
- 2U r^2 e^{2\gamma} du d\Theta + r^2 \big( e^{2\gamma} d\Theta^2+ e^{-2\gamma}
\sin^2\!\Theta\, d\Phi^2 \big) \cma
\end{align}
where
\begin{align}\notag\\[-3em]\notag
\gamma & = \, \frac{c}{r} + {O}(r^{-3}) \\
\beta & = \, -\frac{c^2}{4r^2} + {O}(r^{-3}) \\\notag U & = \, -(c_{,\Theta} + 2c \cot
\Theta ) \frac{1}{r^2} + (3c\, c_{,\Theta} + 4c^2 \cot\Theta) \frac{1}{r^3} + {O}(r^{-4})
\\\notag V & = \, r-2M + {O}(r^{-1}) \cma
\end{align}
$M$ is the Bondi mass aspect and the $u-$derivative of $c$ is the news function.
This metric can also be expressed as (see
\cite{tafel_2}(73)) :
\begin{align}\notag\label{bondi2}
\tilde g = -\Big( 1{-}\frac{2M}{r}+{O}(r^{-2}) \Big) du^2 - \big(2+{O}(r^{-2})\big)du\, dr
-  & \big(n^B{}_{A|B} + {O}(r^{-1})      \big) du\,
dx^A \\
& + r^2\big( s_{AB} -\frac{1}{r} n_{AB}  + {O}(r^{-2}) \big) dx^A dx^B \cmma
\end{align}
$s_{AB}$ being the metric of a two-sphere, $M$ the Bondi mass aspect and $n_{AB}$ the news tensor,
related to the news function $c_{,u}$.

With the general formula for $M$ and $c_{,u}$ determined, we will then investigate their behaviour in various
limits of the C-metric. In section \ref{SmallMassLim} the small mass limit will be computed,
leading to the Minkowski spacetime with the black hole reduced to a
uniformly accelerated particle,
and in section \ref{SchwLim} the small acceleration limit leading to the Schwarzschild limit will be examined.
Then, the large and small Bondi time limits will be investigated in section \ref{ulimits}
for the general case of the C-metric.  Then
in the last section, \ref{genpolarlimit}, the qualitative behaviour of the Bondi mass aspect $M$
and the news function $c_{,u}$ that was observed in the small mass and small acceleration limits will be
investigated, and compared with the general case.

\section{Conformal factor}
\label{Cmetric-cf}

\begin{figure}[h]

 \begin{minipage}{0.48\linewidth}
  \begin{center}
  \includegraphics[height=6.2cm]{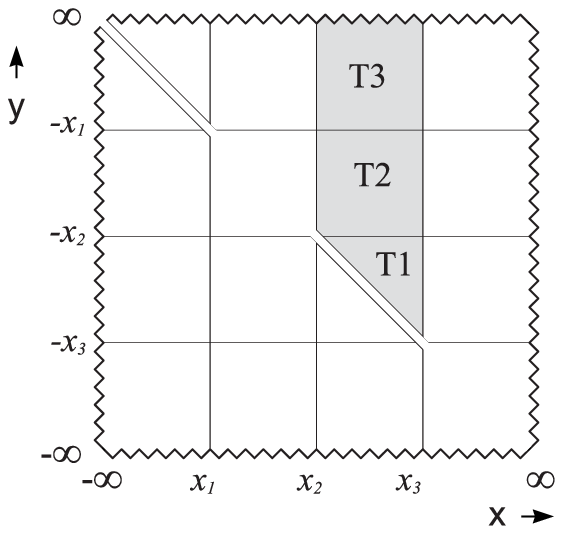}
  \caption{C-metric coordinate ranges.} \label{CMsquare}\notag
  \end{center}
\end{minipage}
\begin{minipage}{0.48\linewidth}

\begin{center}
\includegraphics[height=4.2cm]{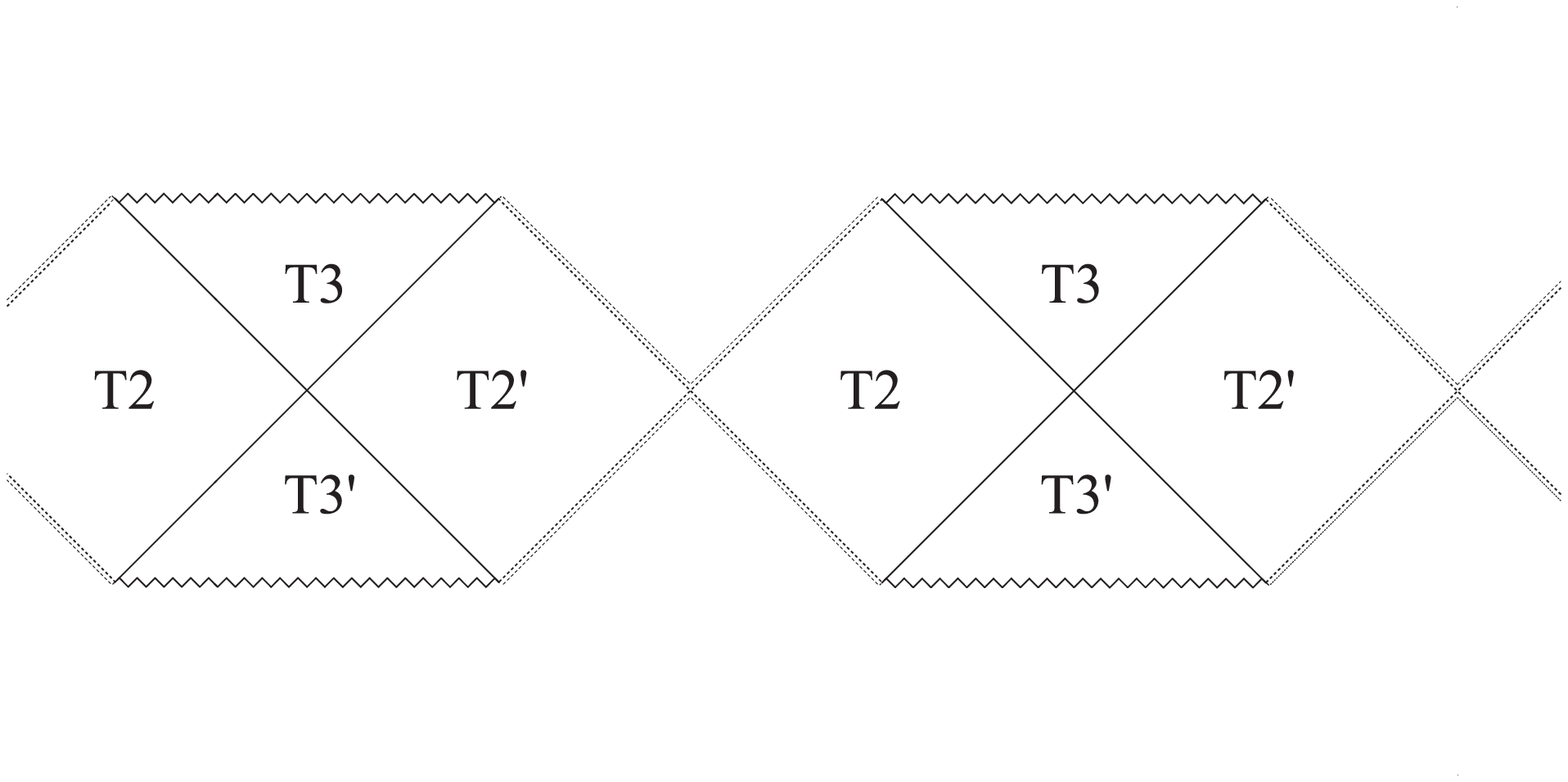}
\caption{Conformal diagram of the $x{=}x_2$ slice.} \label{x2slice}\notag
\end{center}

\begin{center}
\includegraphics[height=4.2cm]{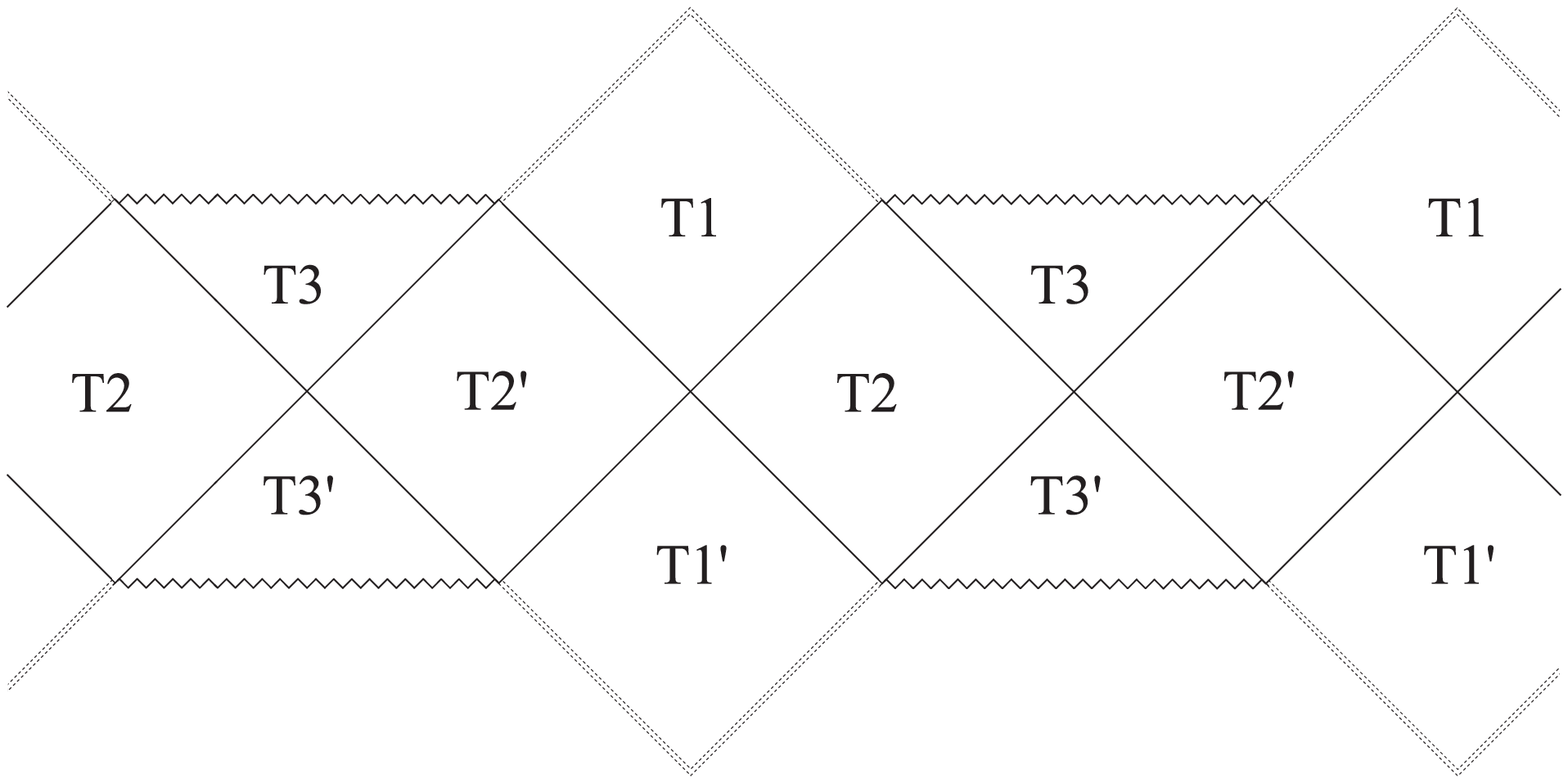}
\caption{Conformal diagram of the $x=x_3$ slice.} \label{x3slice}\notag
\end{center}

\end{minipage}

\end{figure}

\begin{figure}[h]
\begin{center}
\includegraphics[height=7.5cm]{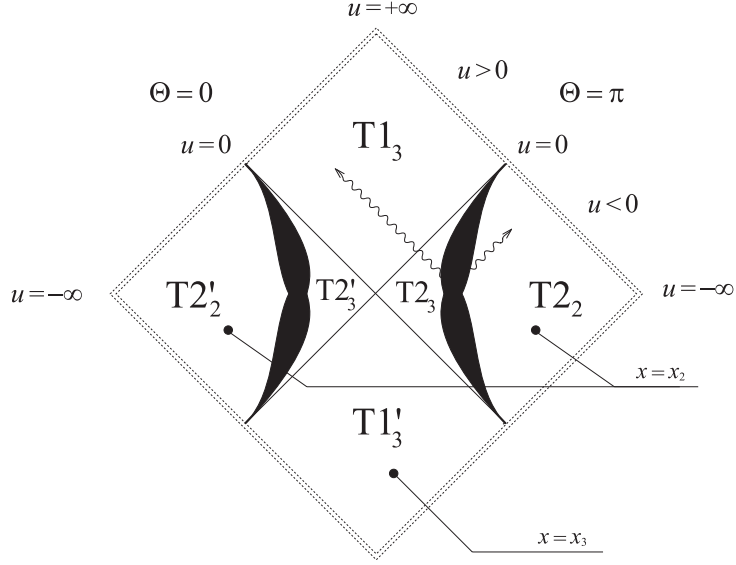}
\caption{A schematic combination of the conformal slices $x{=}x_{2,3}$ into a single diagram, using a
small $mA$ flat space intuition. The interior structure
of the black hole (blocks T3) and the other asymptotic regions are not shown, being
located under the black hole horizon in the black area. The black hole horizon still hints its bifurcation structure,
as seen on figure \ref{x2slice} and \ref{x3slice}.
On $\scri^+$, the angular coordinate $\Theta$ and the Bondi time
$u \hat= \hat u$ are plotted.} \label{slices}\notag
\end{center}
\end{figure}

The C-metric is usually given in the following standard form, which is also most useful
for our calculations:
\begin{equation}\label{CM}
\tilde g=\frac{1}{A^2(x+y)^2}\Big[-F(y) dt^2 + \frac{dy^2}{F(y)} + \frac{dx^2}{G(x)} +
G(x) {\K}^2 d\varphi^2 \Big] \cma
\end{equation}
where $\K$ is the conicity parameter which determines the physical conicity (the ratio of circumference
around a circle to $2\pi$ times the radius, see (\ref{congen}), (\ref{CMcon})) on the axis of symmetry.  The functions $ -F(-z)=G(z)$ are, in
 the vacuum case, cubic polynomials, usually parametrized in one of the two following forms:
\begin{subequations}\label{GPar}
 \begin{align}\label{GPar-a}
  G(x)= & -a\,(x-x_1)(x-x_2)(x-x_3) \\   \label{GPar-b}
  \equiv & \ 1-x^2-2mAx^3 \cma
 \end{align}
\end{subequations}
where the first form  is useful for explicit computations, while the latter is well adapted
for taking limits, as will be studied in sections \ref{SmallMassLim} and \ref{SchwLim}.
There is also no loss of generality associated with the particular gauge (\ref{GPar-b}); one
can always perform a coordinate transformation which translates (\ref{GPar-a}) into
(\ref{GPar-b})\cite{self_1}.
The C-metric actually describes four distinct spacetimes (\cite{self_1}, \cite{PaP}, \cite{krt})
specified by the range of $x$ and $y$ coordinates. Our interest is in the most physically reasonable
one, interpretable as black holes accelerated along a segment of the axis, with a conical singularity;
it is
defined by considering $x \in (x_2,x_3)$ and $y \in (-x,\infty)$, where $x_2$ and
$x_3$ are the two largest roots of $G(x)$, see figure (\ref{CMsquare}). The global conformal
extension of this spacetime is schematically depicted on figures (\ref{x2slice}) and (\ref{x3slice}).
From this and also from a combined figure (\ref{slices}), we can immediately
see that the $x=x_3$ slice corresponds to the inner axial segment between the accelerating
black holes, and the slice $x=x_2$ to the external part of the axis of the symmetry.

Since infinity is located at $x=-y$, and since we would also like to use a null (or retarded) time coordinate,
it is convenient to change the metric into the following form:
\begin{equation}\label{CMnull}
\tilde g=\frac{1}{\rho^2}\Big[{-}F dw^2 + \frac{2}{A}\, dw d\rho -2dw dx + \frac{dx^2}{G} +
G \K^2d\varphi^2 \Big] \cma
\end{equation}
by the transformation $\rho=A(x+y)$, $w=t+\int\! dy/F(y)$.

We now begin our sequence of choices for the switch to an unphysical, conformally equivalent
metric $g=\Omega^2 \tilde g$, such that all of the
properties (a-e), above, are satisfied. The first, quite natural choice is
$\Omega\equiv\Omega_0=\rho=A(x+y)$. Therefore the unphysical metric is now given by the following:
\begin{equation}
g= {-}F dw^2 + \frac{2}{A}\, dw d\rho -2dw dx + \frac{dx^2}{G} +
G \K^2 d\varphi^2  \cma
\end{equation}
with its contravariant form as follows:
\begin{equation}
g^{-1}= 2A\,\partial_w \partial_\rho + A^2(F{+}G)\, \partial_\rho^2+2AG\,\partial_\rho \partial_x +
G\partial_x^2+\frac{1}{G\K^2} \,\partial_\varphi^2\cmb
\end{equation}

However, the condition (d) above
   is not satisfied; i.e., the pullback\footnote{Using expansion
$F(y)=F(\frac{\rho}{A}{-}x)=-G(x{-}\frac{\rho}{A})={-}G(x)+G'(x)\frac{\rho}{A} + {O}\!\left(
\frac{\rho^2}{A^4}\right)$. Also note that in our coordinates, pullback of a form simply means to
 disregard all components containing $d\rho$.} of this metric on  $\scri^+$, namely
\begin{equation}
\phi^*g \equiv g_2= G dw^2  -2dw dx + \frac{dx^2}{G} + G\K^2d\varphi^2 = \left(\sqrt{G}\,dw -
{\frac{dx}{\sqrt{G}}}\right)^2 + G\K^2d\varphi^2   \cma
\end{equation}
is not the metric $g_S$ of a unit 2-sphere. To correct this, we improve our $\Omega_0$ by
multiplying it with another factor, so that the resulting new choice, namely
$\Omega_0 \Omega_S$ will satisfy this condition, which requires
\begin{equation}\label{2sph}
\Omega_S^2 \Big[ G dw^2  -2dw dx + \frac{dx^2}{G} + G\K^2d\varphi^2 \Big] =
\frac{4\,d\xi d\bar\xi}{(1+\xi\bar\xi)^2}= d\Theta^2 + \sin^2\!\Theta \,d\Phi^2
\end{equation}
presented for both stereographic and standard angular coordinates,
which are related as:
\begin{align}\label{stereog}
\xi = e^{i\Phi} \tan\!  \frac{\Theta}{2} \cma \qquad \bar\xi = e^{-i\Phi} \tan\!
\frac{\Theta}{2} \cmmb
\end{align}

If we wish to relate the C-metric coordinates with the asymptotic angular coordinates as
simply as possible, we choose
$\varphi$=$\Phi$.  (If we need some other particular coordinates on $\scri^+$ later, we can always use
some BMS transformation later to transform to them.) Comparing first the coefficients at $d\phi^2$ and $d\Phi^2$, and then
the rest, we obtain :
\begin{align}\notag\label{Sdef}
\Omega_S^2 = \frac{\sin^2 \!\Theta}{G\K^2} \cma \qquad
\frac{\K^2d\Theta^2}{\sin^2\!\Theta}=dw^2-\frac{2dw\, dx}{G} +& \frac{dx^2}{G^2} = \Big (
dw-\frac{dx}{G} \Big)^2   \quad  \\[4pt]\notag
 \Rightarrow \text{we can choose:} \quad \frac{\K d\Theta}{\sin\!\Theta}&=dw-\frac{dx}{G}
 \equiv ds \equiv \K d\S \\[4pt]
 &\Rightarrow \K \S=s= w - \int\frac{dx}{G} \equiv w{-}Gi(x)
\end{align}
Integrating this, we obtain
\begin{equation}\label{OmS}
\S=\ln \bigg( \frac{1{-}\cos\!\Theta}{\sin\!\Theta}\bigg) \ \Rightarrow \quad e^{\ds \S}=
\tan\frac{\Theta}{2} \ \Rightarrow \quad \ch\! \S = \frac{1}{\sin\!\Theta} \cma \ \ \sh\!
\S = -\cot\!\Theta \ \cma \ \ \Omega_S=\frac{1}{ \csh \!\S \, \K G^{1/2} }
\end{equation}
Using the relation\footnote{It is also interesting to note that on $\scri^+$ the $s$ coordinate behaves as
$ds=dt{+}\frac{dy}{F}{-}\frac{dx}{G}=dt{+}{O}(\rho) \ {\hat=\ } dt$.}
 between $S$ and $\Theta$, we can express the stereographic coordinates as
follows:
\begin{align}
\xi = e^{i\Phi} e^S \cma \qquad \bar\xi = e^{-i\Phi} e^S \cmb
\end{align}

As already noted there is a sequential approach to modifying the conformal factor and the
coordinates so as to obtain a form that presents clearly that the metric has the desired
asymptotic properties.  Following Tafel's approach\cite{tafel_1}, we next need to satisfy his
equations (34) and (38), which, in our $(-+++)$ signature, take the form:
\begin{equation}\label{calcon}
\Omega^{-2} \Omega^{|\mu} \Omega_{|\mu} \hat=\ 1 \cma
\quad \Omega^{-1}\, \Omega^{|\mu}{}_{\mu} \hat=\ 2 \cma
\end{equation}
The first may be understood as a calibration condition, which transforms the $\Omega$ into
a form suitable for direct insertion into general formulas for the news function and mass
aspect. The latter equation is related to the condition on the determinant of the resulting Bondi-Sachs
 coordinates (see (7) and (38) of \cite{tafel_1}). If the conditions (a)--(e)
 are satisfied, then the latter follows from the former.

Our $\Omega'\equiv\Omega_0\Omega_S$ unfortunately doesn't yet satisfy (\ref{calcon}), and
therefore needs to be modified further.
To get such an $\Omega$ we correct $\Omega'$ using the gauge freedoms given in equation (35) of
 \cite{tafel_1} and equation (52) of  \cite{tafel_2}, which we repeat below:
\begin{equation}\label{Om_final}
\Omega=\Omega'+\eta\Omega'^2 \cma \qquad \eta \eqhat  -\frac{1}{2} \hat u^{|\mu}{}_{\mu} -
\Omega'^{-1}\,(1 {-} \Omega'^{|\nu} \hat u_{|\nu} )
\end{equation}
The coordinate $\hat u$ is an asymptotic Bondi time,
which coincides with the Bondi time coordinate on
$\scri^+$. The function $\hat u$ can be obtained from equation (16) of  \cite{tafel_1}
and equation (48) of \cite{tafel_2}, which for
our $\Omega'$ reads :
\begin{equation}\label{uhatEq}
\Omega'^{|\mu} \partial_\mu {=} \ \partial_{\hat u} \hat=\ \partial_u \cma \quad
\Omega'^{|\mu} \hat u_{\mu}\ \hat = \ {}1 \cmmb
\end{equation}
Substituting $\Omega'=\Omega_0\Omega_S$ leads directly to a differential relation:
\begin{equation}
A\,\Omega_S^{-1} (G\partial_x + \partial_w) = \partial_{\hat u} \cma
\end{equation}
which has as solution the following:
\begin{equation}\label{uhat}
\hat u= \frac{1}{A\K \csh \! \S} \int \!\! \frac{dx}{G^{3/2}} \ + \, \alpha(\xi,\bar\xi)
\equiv \frac{1}{A\K \csh \! \S} \, \Gj(x)+ \, \alpha(\xi,\bar\xi) \cma
\end{equation}
where we have denoted the integral of $G^{-3/2}$ as $\Gj$.
The explicit form and some other properties of this function can be found in the appendix, see (\ref{gj-intexplicit-C2}),
(\ref{gj-intexplicit-C3}) and also figure (\ref{FigGGj}).
The quantity $\alpha(x^A)$ includes the integration constant of the
function $\Gj$.  It is an arbitrary function of the asymptotic angular
coordinates only; i.e., it does not depend on the Bondi time $\hat u$,
and corresponds to the supertranslations contained in the BMS group of $\scri^+$ coordinate
transformations.

This result for $\hat u$ together with  (\ref{Om_final})
allows us to obtain\footnote{\label{rems} Interestingly,
 had we used the coordinates $(w,\rho, \Theta,\phi) $,
  the formula (\ref{Om_final}) for $\hat u $ would have
 simplified significantly, giving $\eta = -\frac{1}{2}\triangle \hat u$,
  where $\triangle$ is the Laplace operator on the $2-$sphere
  and the derivatives are to be  taken while holding $w$
  constant. This is probably because in the $(w,\rho, \Theta,\phi) $
   coordinates the angular part of the metric is exactly that of the 2-sphere, and so
   are the corresponding $\Gamma^{A}{}_{B}{}_{C}$. Effectively, what
   happens is that the $\Omega^{-1}\,(1 {-} \Omega^{|\nu} \hat u_{|\nu} )$ term
   precisely cancels with $-\frac{1}{2}( \hat u^{|w}{}_{w} + \hat u^{|\rho}{}_{\rho})$.
   }
 $\eta$ and therefore the final conformal factor:
\begin{align}\label{eta2}\notag
\Omega_F=\ &\Omega_0\Omega_S \,(1{+} \eta\, \Omega_0\Omega_S )\cmma\quad
\Omega_0=\rho \cma \quad \Omega_S=\frac{1}{\K \csh\!\S \, \sqrt{G}}  \cmma
 \\[0.5em]
\eta= \frac{G\!j}{2A\K\ch\! \S}&\,(1{-}\sh^2\!\! \S) +\frac{\K G'}{4A\sqrt{G}}\,\ch\!\S  -\frac{\sh \!\S}{A\sqrt{G}}
- \frac{1}{2}\triangle \alpha
 \cma
\end{align}
where $\Lapl$ is the Laplace operator on the $2-$sphere:
$\Lapl =(1+\xi\bar \xi)^2\pdiff{{\xi}}\pdiff{{\bar\xi}}
= \pdiff{\Theta} \pdiff{\Theta} {+} \cot\!\Theta \,\pdiff{\Theta} {+}
\sin^{\!-2}\!\Theta \,\pdiff{\phi} \pdiff{\phi} {}$ .

 Using the $\Omega {\equiv} \Omega_F$ in the transition $\tilde g \rightarrow g{=}\Omega^2 \tilde g$ now finally ensures that
 both conditions
(\ref{calcon}) are satisfied. The dependence on $\alpha$ is also in agreement with the
transformation properties of $\eta$ under the supertranslations, see \cite{tafel_2}(32).
From now on, we will simply denote $\Omega_F$ as $\Omega$.

\section{Mass aspect and the news function}

The final conformal factor $\Omega\, {\equiv}\, \Omega_F$ will now be used, according to the
scheme outlined at the end of section \ref{introduction}, to obtain the
Bondi mass aspect $M$ and the news function $c_{,u}$.
According to the procedure described in Sec.3 of \cite{tafel_2}, we start with the modified mass aspect $\hat
M$. In our case the explicit formula \cite{tafel_2}(44) for the modified mass aspect
 and its change under the supertranslations subgroup of the BMS transformations yields, for the final conformal factor $\Omega$ :
\begin{align}\notag\label{MH}
\hat M \hat = \ \Omega^{-1} & (1-2\,\Omega^{-2} \Omega^{|\mu}
\Omega_{|\mu}+\frac{1}{2}\Omega^{-1}\Omega^{|\mu}{}_\mu ) -{\txts \frac{1}{4}} \, (\Lapl{+}2)\Lapl \alpha   \\
&= \frac{\csh^3\!\S} {4A\K\sqrt{G}}\, \Big[ \frac{\K^4 G'^3}{8} {-} \K^2G' {-} \frac{1}{4}\K^4 G G' G''
{+} \frac{1}{6}\,\K^4G^2 G''' {-} \sqrt{G} \,G\!j \Big]   -{\txts \frac{1}{4}} \, (\Lapl{+}2)\Lapl \alpha
\end{align}
and its $u-$derivative
\begin{align}\label{MHu}
\hat M_{,u} \ \hat = \ {-}\frac{1}{4} \csh\!\S^4 \Big[ 1+ \frac{1}{2}\K^2 GG'' - \frac{1}{4}\K^2G'^2
\Big]^2 \cmb
\end{align}
The behaviour under the supertranslation $\alpha$ simply follows from
$\alpha$-dependence of the expression (\ref{eta2}) used in the $\Omega_F$, in accord with the general formula \cite{tafel_2}(41).

To find the news function $c_{,u}$, we first compute the news tensor as decried in  \cite{tafel_2} (67) or (61):
\begin{align}\notag\label{newsdef}
n& =-\varphi^*(L_vg)=-2\varphi^*(u_{|\mu\nu} dx^{\nu}dx^{\mu} ) \cma \qquad v\equiv u^{|\nu}\partial_\nu \\
\dot n & = -\varphi^*(\Omega^{-1} L_vg) = -2\varphi^* (\Omega^{-1}\Omega_{|\mu\nu}
dx^{\mu} dx^{\nu} )
\end{align}
using also \cite{tafel_2}(69), which in our case reads:
\begin{align}\notag
n'_{AB}=n_{AB}-2\, \alpha_{|AB} + \Lapl \alpha\, {g_S}_{AB} \cmb
\end{align}
 This leads\footnote{The same simplification can occur as in the case of $\eta$, see remark (\ref{rems}).
 Also, the $\alpha$ correction vanishes identically for $n_{\xi\bar\xi}$, ensuring the
 tracelessness of $n_{AB}$.}
  to the news tensor $n_{AB}$ being :
\begin{align}\label{newstensor}
n_{\xi\xi} \ \hat=\ {-}\frac{1}{\xi\xi}\, \frac{1}{2A \K \csh\!\S} \Big(
G\!j+\frac{\K^2G'}{2\sqrt{G}} \Big) -2\, \alpha_{|\xi\xi}
\cma \qquad n_{\bar \xi \bar \xi}=\bar n_{\xi\xi} \cmma
\quad n_{ \xi \bar \xi}=0
\end{align}
which is consistent with (\ref{MHu}), seen by using \cite{tafel_2}(66).

Now we can directly compare the angular parts of the Bondi metric forms (\ref{BondiMetric}) and (\ref{bondi2}) :

\begin{align}\label{n_c_rel}\notag\\[-2.5em]
\xi^2\,(\sin\Theta)^{-2}\,n_{\xi\xi}+ c.c.
=n_{\Theta\Theta}={-}(\sin\Theta)^{-2}\,n_{\Phi\Phi} = -2c \cma
\end{align}
leading to :
\begin{align}\label{c_u}\notag\\[-2em]\notag
\quad c  =\ &  {-}\frac{1}{2A\K\sin \Theta} \,\Big[ \Gj+
\frac{\K^2G'}{2\sqrt{G}} \Big] + {\txts \frac{1}{2}} \alpha_{|\Theta\Theta} {-}{\txts \frac{1}{2}} \sin^{\!-2}\!\Theta \,\, \alpha_{|\Phi\Phi}
\cma \\[0.5em]
c_{,u}  =\ &  {-}\frac{1}{2\sin^2 \! \Theta} \,
\Big[ 1{+}\frac{1}{2}\K^2 GG'' {-} \frac{1}{4}\K^2 G'^2 \Big] =
 {-}\frac{\K^2}{2\sin^2 \! \Theta}\, \Big[ 1{+} \frac{1}{2}\ GG'' {-} \frac{1}{4} G'^2 \Big]+ \frac{\K^2{-}1}{2\sin^2\Theta} \cmb
\end{align}

Next we will use our result for $c$ to obtain the Bondi mass aspect $M$ from the reduced mass aspect $\hat M$, using the formula \cite{tafel_2}(72):
\begin{align}\label{bondiM}
M=\hat M - \frac{1}{4}n^{AB}{}_{|AB} = \hat M +  \frac{1}{2} ( c_{,\Theta\Theta}+3c_{,\Theta} \cot\Theta -
2c -  c_{,\Phi \Phi}\,\sin^{\!-2}\Theta)\cma
\end{align}
which we have expressed in terms of the news function $c(u,\Theta, \phi)$, using the relation (\ref{n_c_rel}).

Together with one of the two Bondi's supplementary conditions\footnote{A consequence of
the vacuum Einstein equation, namely $R_{00}=0$, in coordinates (\ref{BondiMetric}), see \cite{Bondi} for further discussion.},
assuming\footnote{This is
just a technical simplification, since the version of Bondi's analysis covered in textbooks typically assumes
axial
symmetry. Of course, the C-metric case covered here is also axially symmetric.}
 $c{=}c(u,\Theta)$  only, i.e., it is axially symmetric,
\begin{align}\label{BondiSup}\notag
M_{,u} = -c^2_{,u} + \frac{1}{2} ( c_{,\Theta\Theta}+3c_{,\Theta} \cot\Theta - 2c)_{,u}
\cmma
\end{align}
we obtain an interesting relation:
\begin{align}
M_{,u}= -c^2_{,u} - \frac{1}{4}\big(n^{AB}{}_{|AB}\big)_{,u} \ \Rightarrow \hat M_{,u} =
-c^2_{,u} \cma
\end{align}
in agreement with (\ref{MHu}) and \cite{tafel_2}(66).

Substituting (\ref{c_u}) into (\ref{bondiM}), and using the partial derivatives (\ref{parcM}), we finally
obtain the value for $M$ itself:
\begin{align}
\notag\label{bondiM-CM}
 M=  \frac{1}{4\sin^3 \! \Theta} \, \Big[ 1{+}
{\txts\frac{1}{2}}\K^2 GG''  {-} & {\txts\frac{1}{4}}\K^2 G'^2 \Big]\Big(
\alpha_{,\Theta\Theta}  \sin\!\Theta {-} \alpha_{,\Theta} \cos\!\Theta   {-}
\frac{\Gj}{A\K}\Big) - \frac{A\K^3G^{\frac{5}{2}}
G'''}{8\sin^3\!\Theta}\Big(\alpha_{,\Theta}{+} \cos\!\Theta
\frac{\Gj}{A\K} \Big)^2 \\
&+
\frac{1}{8A\!\sqrt{G} \sin^3\!\Theta} \Big[ \txts
\frac{1}{4}\K^3G'^3{-}\K G'{-}\frac{1}{2}\K^3GG'G''{+}
\frac{1}{3}\K^3G^2G''' \Big] \cmmb
\end{align}

It is interesting to notice the absence of a pure $\alpha$ correction term: after putting the $\alpha$ terms from
(\ref{newstensor}), (\ref{MH}) together into the definition (\ref{bondiM}) one would expect to
obtain
\begin{align}\notag
 -\txts \frac{1}{2}  \Delta\Delta \alpha - \frac{1}{2} \Delta\alpha + \frac{1}{2}\alpha_{|AB}{}^{AB} \,  \cma
\end{align}
in (\ref{bondiM-CM}), but surprisingly, due to the simplicity of the Riemann tensor in two dimensions, this correction
term is identically zero for any $\alpha(\Theta,\phi)$. Therefore the mass aspect $M(\hat u,\Theta,\phi)$ transforms
itself under the supertranslations
$\hat u \rightarrow \hat u' {=} \hat u{+}\alpha(\Theta,\phi)$ only due to the explicit change of
$\hat u$, and due to the change of the partial derivatives\footnote{
In our case this  manifested itself via (\ref{parcM}). Had
we been able to express the $n_{AB}$ in terms of $\hat u$, $\Theta$ and $\phi$ only, the $\alpha$
would have emerged as a result of $ \Diff{\hat u}{\Theta'} {=} {-}\alpha_{,\Theta}$ and $ \Diff{\hat u}{\Phi'}
{=}{-}\alpha_{,\Phi}$ in the transformation $\hat u' {=} \hat u {+}\alpha$, $\Theta'{=}\Theta$ and
$\Phi'{=}\Phi$.} in
$\frac{1}{4}n_{AB}{}^{|AB}$.

Last but not least, there is an interesting 'alternative' result for $M$. If we had used $\Diff{x}{\Theta} {}_{|w=\text{const.}} $
instead of $\Diff{x}{\Theta} {}_{|\hat u=\text{const.}} $ in (\ref{bondiM}), we would have
ended with a much more simple expression:
\begin{align}\label{bondiM-CMw}\notag
M^{(w)}= {-}\frac{\K^3  G^{\frac{3}{2}}
G'''}{12A\sin^3 \! \Theta}=m \K^3  G^{\frac{3}{2}} \csh^3\!\S = m\Omega_S^{-3} \cmb
\end{align}
However its interpretation is not clear, since $w$ is not a Bondi time\footnote{On the other hand
$w$ becomes the Bondi time asymptotically in the Schwarzschild limit,
see section \ref{SchwLim}. It might be possible to regard this mass aspect as a mass aspect
for a stationary observer adapted to (at rest with respect to) the accelerated black
hole, who becomes an asymptotical Schwarzschild observer in the $A \rightarrow 0$ limit.}.

\section{Small mass limit}

\label{SmallMassLim}
In this section we will find the limit of the news function $c_{,u}$ and the mass aspect $M$ for small mass $m$.
Substituting $G(x)$ in the form of (\ref{GPar-b}) into (\ref{c_u}), the news function $c_{,u}$ reduces as follows:
\begin{align}
c_{,u}= \frac{\K^2{-}1}{2\sin^2 \Theta} + \frac{A\K^2mx}{\sin^2 \Theta} \big[ x^2{-}3
\big] + O(mA)
\end{align}
To obtain this in terms of the asymptotic Bondi coordinates only, we express $\Gj(x)$ from (\ref{uhat}).
Then, using the expansion (\ref{mAexpansion-x}) together with (\ref{kappaExp}), we arrive to:
\begin{align}\label{newsm0}
c_{,u}= \frac{\kappa^2{-}1}{2\sin^2 \Theta} \pm \frac{2\kappa^2 \!A}{\sin^2 \Theta} \,m
 + \frac{\hat u A^2 \kappa^3}{\sin^2 \Theta}\,.\,
 \frac{2\hat u^2 A^2 \kappa^2{+}3\sin^2\Theta}{(\hat u^2 A^2 \kappa^2{+}
 \sin^2\Theta)^{3/2}} \, m
 + O(m^2)
\end{align}
where $\kappa$ denotes the physical conicity only between particles (i.e. $\kappa_{ext}{=}1$) for the '$-$' sign, or outside of
the particles (i.e. $\kappa_{in}{=}1$) for the '$+$' sign respectively.
This is in agreement with the special case of $\kappa_{ext}{=}1$ presented in \cite{PaP}. It is closely related to
news functions given in \cite{bicak_1}, \cite{BiBo} and \cite{BiRo}. Note that the first term corresponds precisely to the
news function of an infinite cosmic string; see equation (7) in \cite{BiAstr} and (25) in \cite{bicak_3}.

\begin{figure}[h]
 \begin{center}
  \includegraphics[height=7.5cm]{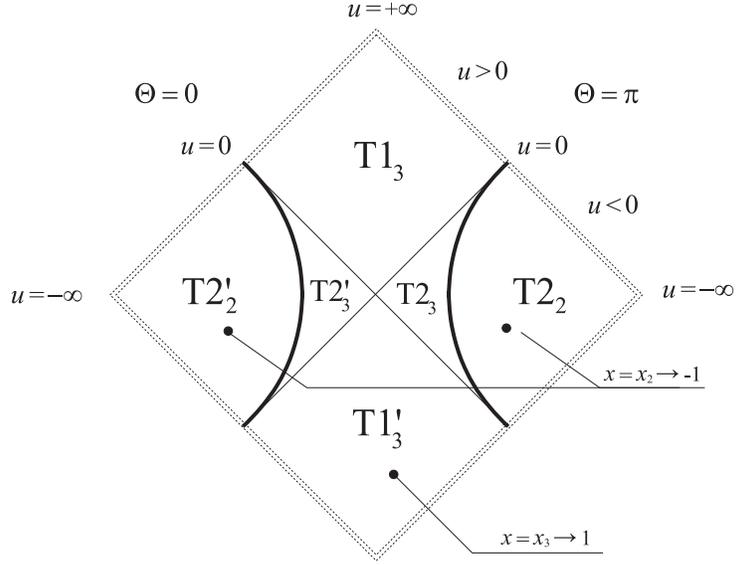}
  \caption{Conformal diagram depicting the Minkowski limit $m{\rightarrow}0$.
   On the $\scri^+$, the Bondi angular coordinate $\Theta$ is
   shown, together with the Bondi null time $u$. The black hole is reduced to a uniformly accelerated pointlike particle.}
  \label{FigMinkwLim}\notag
 \end{center}
\end{figure}

It is of interest to see how exactly are the possible singularities of the function $c_{,u}$
distributed on $\scri^+$. First, with the help of our equation (\ref{uhat}) and the properties of
the $\Gj(x)$ function (see figure \ref{FigGGj}), we realize the correspondence between the roots $x=x_{2,3}$ and poles
$\Theta=0,\pi$:
\begin{align}\label{poles-m}
\hat u&>0 :  \ \Big\{
    \begin{array}{lcl}
        \Theta{=}0 & \leftrightarrow &   x{=}x_3  \\[0.55em]
        \Theta{=}\pi & \leftrightarrow  & x{=}x_3
    \end{array}\cmma \qquad
\hat u<0: \ \Big\{
   \begin{array}{lcl}
        \Theta{=}0 &\leftrightarrow & x{=}x_2  \\[0.55em]
        \Theta{=}\pi &\leftrightarrow  & x{=}x_2
    \end{array} \cmmb
\end{align}
This is illustrated in figure \ref{FigMinkwLim}.

It is obvious that singular behaviour can only occur on the
axis, i.e. for $\Theta=0,\pi$. Expanding around the poles yields the following:
\begin{align}\notag\label{c_u-A0}
\kappa_{ext}&=1: \quad \Bigg\{
\begin{array}{l}
\ds c_{,u}=2mA\,(\sign  \hat u{+}1) \frac{1}{\Theta^2 }+ \frac{2}{3}mA \,(\sign \hat u{+}1) +O(\Theta^2) \\[0.55em]
\ds c_{,u}=2mA\,(\sign \hat u{+}1) \frac{1}{(\pi{-}\Theta)^2} + \frac{2}{3}mA\,(\sign \hat u{+}1)
+O\big((\pi{-}\Theta)^2\big) \cmma
\end{array} \\[1em]
\kappa_{in}&=1: \quad \Bigg\{
\begin{array}{l}
\ds c_{,u}=2mA\,(\sign \hat u{-}1) \frac{1}{\Theta^2 }+ \frac{2}{3}mA \,(\sign \hat u{-}1) +O(\Theta^2) \\[0.55em]
\ds c_{,u}=2mA\,(\sign \hat u{-}1) \frac{1}{(\pi{-}\Theta)^2} + \frac{2}{3}mA\,(\sign \hat u{-}1)
+O\big((\pi{-}\Theta)^2\big) \cmmb
\end{array}
\end{align}
Therefore by choosing the conical singularity to exist only between the particles, it
appears on $\scri^+$ only for $u>0$, that is above the acceleration Cauchy horizon $y=-x_1$,
and vice versa (see figure \ref{CMsquare} and \ref{slices}).
It is interesting that, as shown in section \ref{genpolarlimit}, this situation occurs not
only in the small $m$ limit, but also persists in the full C-metric.

Also, perhaps not very surprisingly, this behaviour translates into the properties of the
Bondi mass aspect $M$. Using the same approach as in the case of the news function, application of (\ref{uhat}),
(\ref{mAexpansion-x}) and (\ref{kappaExp}) for (\ref{bondiM-CM}), while assuming  $\alpha{=}0$, yields:
\begin{align}\notag\label{Mmlim}
M(\hat u, \Theta)={} & \frac{m}{2U^5 \sin^3\!\Theta}  \,\big(u'^2(3\cos^2\!\Theta{-}1) - 1 \big) -
     \frac{m^2Au'}{U^6 \sin^3\!\Theta} \, \Big( 8u'^6 + 24 u'^4+ u'^2(15\cos^2\!\Theta{ +}20) +4
     \Big)  \\\notag
          & \quad \quad \pm \frac{m^2A}{U^7 \sin^3\!\Theta} \, \Big( 8u'^8+ 28u'^6 +u'^4(6\cos^2\!\Theta
          {+}34)-u'^2(9\cos^2\Theta {-}19) +5 \Big) \ +\ O(m^3) \cma    \\
        & \hskip14em \text{where } \quad U=\sqrt{ \vphantom{{A^3{}}^3} u'^2+1} \cma \quad u'= \frac{\hat uA}{\sin\Theta} \cmma
\end{align}
and the '$-$' sign denotes the case with the axis regular outside of the particles, i.e.,
$\kappa_{ext}{=}1$, while the '$+$' sign assumes the axis being regular between the particles,
i.e.,
$\kappa_{in} {=} 1$.

To investigate the integrability of our mass aspect $M$, we expand (\ref{Mmlim})
near the poles $\Theta{=} 0$ ($x{=}x_3$) and $\Theta{=}\pi$ ($x{=}x_2$), obtaining:
\begin{align}\notag\label{m_m0}
\kappa_{ext}&=1: \quad \Bigg\{
\begin{array}{l}
\ds M=-8m^2\!A^2\hat u\,(1{+}\sign \hat u) \frac{1}{\Theta^4 }  -  \frac{16}{3}m^2\!A^2\hat u \,(1{+}\sign \hat u) \frac{1}{\Theta^2 } +O(1) \\[0.55em]
\ds M=-8m^2\!A^2\hat u\,(1{+}\sign \hat u) \frac{1}{(\pi{-}\Theta)^4} -  \frac{16}{3}m^2\!A^2\hat u \,(1{+}\sign \hat u) \frac{1}{(\pi{-}\Theta)^2 }
+O(1) \cmma \\[0.55em]
\end{array} \\[1em]
\kappa_{in}&=1: \quad \Bigg\{
\begin{array}{l}
\ds M=-8m^2\!A^2\hat u\,(1{-}\sign \hat u) \frac{1}{\Theta^4 }  -  \frac{16}{3}m^2\!A^2\hat u \,(1{-}\sign \hat u) \frac{1}{\Theta^2 } +O(1) \\[0.55em]
\ds M=-8m^2\!A^2\hat u\,(1{-}\sign \hat u) \frac{1}{(\pi{-}\Theta)^4} -  \frac{16}{3}m^2\!A^2\hat u \,(1{-}\sign \hat u) \frac{1}{(\pi{-}\Theta)^2 }
+O(1) \cmmb
\end{array}
\end{align}
As expected, the behaviour is qualitatively the same as in the case of $c_{,u}$.

Furthermore, we find the mass change\footnote{The function $M$ is fortunately sufficiently continuous that
the $\hat u$-derivative commutes
with the $m {\rightarrow} 0$ limit.}, $M_{,u}$ to be:
\begin{align}\notag\label{Mumlim}
M_{,u}(\hat u, \Theta)= { } &  {-}\frac{3mu'A}{2U^7 \sin^4\!\Theta}  \,\big(u'^2(3\cos^2\!\Theta{-}1) - 1{-}2\cos^2\!\Theta \big)
      \\\notag
       &  \qquad \qquad - \frac{m^2A^2}{U^8 \sin^4\!\Theta} \, \Big( 8u'^8 +32u'^6 + 15u'^4(4{-}3\cos^2\!\Theta) +5u'^2(9\cos^2\!\Theta{+}8)
     +4 \Big)
       \\[0.1em]\notag
          & \hskip-1.5em \pm \frac{m^2A^2u'}{U^9 \sin^4\!\Theta} \, \Big( 8u'^8{+} 36u'^6 {+}6u'^4(11{-}3\cos^2\!\Theta)
               {+}u'^2(41{+} 69\cos^2\!\Theta ) {-}18\cos^2\!\Theta{+}3 \Big) \ +\ O(m^3) \cma    \\
        & \hskip14em \text{where :} \quad U=\sqrt{ \vphantom{{A^3{}}^3} u'^2+1} \cma \quad u'= \frac{\hat uA}{\sin\Theta} \cmma
\end{align}
again, the '$-$' sign denotes $\kappa_{ext}{=}1$, while the '$+$' occurs for $\kappa_{in}{=}1$.

On first sight, one may expect, according to (\ref{bondiQ}), an inconsistency with (\ref{newsm0}),
because of the non-zero term $O(m)$ here in $M_{,u}$. It is, however, still true that
\mbox{$M_{,u}^{\text{(tot.)}}=-\int \! c_{,u}^2 dS = \int \! M_{,u} dS$} (if it exists),
because the aforementioned term $O(m)$ integrates
to zero. This confirms that the lowest order of mass change is indeed $O(m^2)$, as
suggested by $c_{,u}$, and in analogy with the electromagnetic
radiation case, where the Poynting vector, and therefore the energy
radiated per unit time, is proportional to the charge of the accelerated particle squared.

The expansion near the poles is completely analogous to the case of $M$ and can be simply
obtained by applying the derivative with respect to $\hat u$ to (\ref{c_u-A0}):
\begin{align}\notag\label{m_u-m0}
\kappa_{ext}&=1: \quad \Bigg\{
\begin{array}{l}
\ds M_{,u}=-8m^2\!A^2\,(1{+}\sign \hat u) \frac{1}{\Theta^4 }  -  \frac{16}{3}m^2\!A^2 \,(1{+}\sign \hat u) \frac{1}{\Theta^2 } +O(1) \\[0.55em]
\ds M_{,u}=-8m^2\!A^2\,(1{+}\sign \hat u) \frac{1}{(\pi{-}\Theta)^4} -  \frac{16}{3}m^2\!A^2 \,(1{+}\sign \hat u) \frac{1}{(\pi{-}\Theta)^2 }
+O(1) \cmma \\[0.55em]
\end{array} \\[1em]
\kappa_{in}&=1: \quad \Bigg\{
\begin{array}{l}
\ds M_{,u}=-8m^2\!A^2\,(1{-}\sign \hat u) \frac{1}{\Theta^4 }  -  \frac{16}{3}m^2\!A^2 \,(1{-}\sign \hat u) \frac{1}{\Theta^2 } +O(1) \\[0.55em]
\ds M_{,u}=-8m^2\!A^2\,(1{-}\sign \hat u) \frac{1}{(\pi{-}\Theta)^4} -  \frac{16}{3}m^2\!A^2 \,(1{-}\sign \hat u) \frac{1}{(\pi{-}\Theta)^2 }
+O(1) \cmmb
\end{array}
\end{align}

\section{Schwarzschild limit}

\label{SchwLim}
It is also possible to investigate the other situation, which is the case where $m$ is
kept finite and non-zero while $A \rightarrow 0$. Intuitively, this should lead to a single
static blackhole, i.e., to the Schwarzschild solution. To show that this is indeed true, we
start with the physical metric (\ref{CM}). In order to get the correct limit of
the metric tensor, we have to parametrize the coordinates and the parameters $A$, $m$, $\K$,
characterizing the C-metric solution. Perhaps the most physically plausible way to do this is by using $A$
as a parameter\footnote{In the $m\rightarrow 0$ limit, A is precisely the acceleration of the test
particle which the black hole becomes.}, while holding the horizon area
${\mathcal A}$ and the conicity $\kappa_{ext,in}$
 on one of the axial segments constant. This leads to
(see (\ref{CMcon}), (\ref{mASconstExpansion})) :
\begin{align}\label{SchwLimParamScale}\notag
\kappa_{ext}&=1 : \quad
 \quad \K = 1 + 2m'A + \frac{11}{2}m'^2A^2 - 4m'^3 A^3 + O(A^4) \\\notag
&  \hskip 3.5em \quad m= m' - m'^2 A -\frac{21}{2} m'^3 A^2 + \frac{71}{2} m'^4 A^3 + O(A^4)
\\[.75em]\notag
\kappa_{in}&=1 : \quad
\quad \K   = 1 - 2m'A + \frac{11}{2}m'^2A^2 + 4m'^3 A^3 + O(A^4) \\
&  \hskip 3.5em \quad m = m' + m'^2 A -\frac{21}{2} m'^3 A^2 - \frac{71}{2} m'^4 A^3 + O(A^4) \cma
\end{align}
where the horizon area ${\mathcal A}=16\pi m'^2$.
Together with a simple coordinate rescaling:
\begin{align}\label{SchwLimCoordScale}
\begin{array}{l}
y=y'A^{-1}  \\
 t=t'A  \\
 x=x'
\end{array}
& \Rightarrow \quad
\begin{array}{rcrcrcl}
A^2F&=&-A^2+y'^2-2my'^3 &=& -A^2+F' +O(A) \cmma \quad F' &=& y'^2(1{-}2m'y') \\  G&=&1-x'^2-2mAx'^3 & = & \ \
G'-2mAx'^3 \cmma \quad G' &=& 1{-}x'^2
\end{array}  \cma
\end{align}
 the metric $\tilde g$ takes the following reduced form:
\begin{align}
\label{gSchwLim}
\begin{split}
 \tilde g=\frac{1}{A^2(x{+}y)^2}\Big[-Fdt^2 + \frac{dy^2}{F}{ + }\frac{dx^2}{G} + G {\K}^2
d\varphi^2 \Big]   \hskip 16em \\ = \frac{1}{(Ax'{+}y')^2} \Big[ \big(F'+O(A)\big)dt'^2 +
\frac{dy'^2}{F'{+}O(A)} + \frac{dx'^2}{G'{-}2mAx'^3} +(G'{-}2mAx'^3)\K^2 d\varphi^2 \Big]
 \\ 
 \xrightarrow[A\rightarrow 0]{} \quad
\frac{1}{y'^2} \Big[ -F'dt'^2+ \frac{dy'^2}{F'} + \frac{dx'^2}{G'} + G' d\varphi^2 \Big]
\cma
\end{split}
\end{align}
which is indeed the Schwarzschild metric\footnote{To be more precise, it is (for $\K{\neq}1$) the
Schwarzschild metric with a conical singularity.} as
can be seen using the transformation:
\begin{align}\label{gSchwLimMetric}
\begin{array}{l}
\ds y'=1/R \\
 x'=\cos{\theta}
\end{array}
&& \Rightarrow \quad & \tilde g = { \txts -(1{-}  \frac{ 2M}{ R})} \, dt^2
+\frac{dR^2}{(1{-}\frac{2M}{R})} + R^2( d\theta^2{+} \sin^2\!\theta\, d\varphi^2 ) \cma
\end{align}
where the Schwarzschild mass $M$ equals the mass $m'$, confirming the condition ${\mathcal A}{=}\text{constant}$.

 Now we can proceed to
compute the limits of various asymptotic quantities, i.e. obtain the asymptotic expansion
of the metric in Bondi coordinates. The strategy is to obtain a given quantity $f(\hat u,\Theta)$
 as a series in $A$ with coefficients being a function of the limiting coordinates $\theta$ and $u'$ only.
This is however not as straightforward as the $m\rightarrow0$ case, as we will
see when computing the limit of $\hat u$, which we would like to coincide with the Bondi
time for the Schwarzschild metric. To check whether this can be satisfied, it is useful to
realize the relation between the Schwarzschild Bondi time and the limit of the $w$ coordinate
(see (\ref{CMnull})):
\begin{align}\notag
w= t+\int\! \frac{dx}{F} =At'+\int\! \frac{Ady'}{F'{+}O(A)} = \Big(t'{+}\int\!
\frac{dy'}{F'} \, \Big) A +O(A^2) \equiv w'A + O(A^2)  \\\label{wSchw}
\text{while also: } \quad t'+ \int\!\frac{dy'}{F'} = t'- \int\!\frac{dR}{1-\frac{2M}{R}}
\equiv t'-r^* \equiv u' \cmb
\end{align}
 where $u'$ is precisely the Bondi null time\footnote{This can also be verified by
 comparing the Schwarzschild metric in
$(u,r,\Theta,\phi)$ coordinates with (\ref{bondi2}), or by a direct application of
(\ref{uhatEq}).} on $\scri^+$ of the Schwarzschild metric, with $u'$ constant, being the
future directed null cones, as required. In addition, by a suitable
redefinition\footnote{This, of course, does not spoil the limit \ref{gSchwLim}.} of $t'
\rightarrow t'+O(A)$, we can make the $w' \leftrightarrow u' $ correspondence exact,
i.e., $w'=u'$.

The idea is now to use (\ref{Sdef}) to express $x$ as a function of $\Theta$ and $w'$,
and insert this into (\ref{uhat}) to obtain relation between $\hat u$ and $u'$ as a function
of $\Theta$ only. The terms lower than $O(A)$ of this function will then be the needed
supertranslation.

\begin{figure}[h]
 \begin{minipage}{0.47\linewidth}
  \begin{center}
    \includegraphics[height=7.5cm]{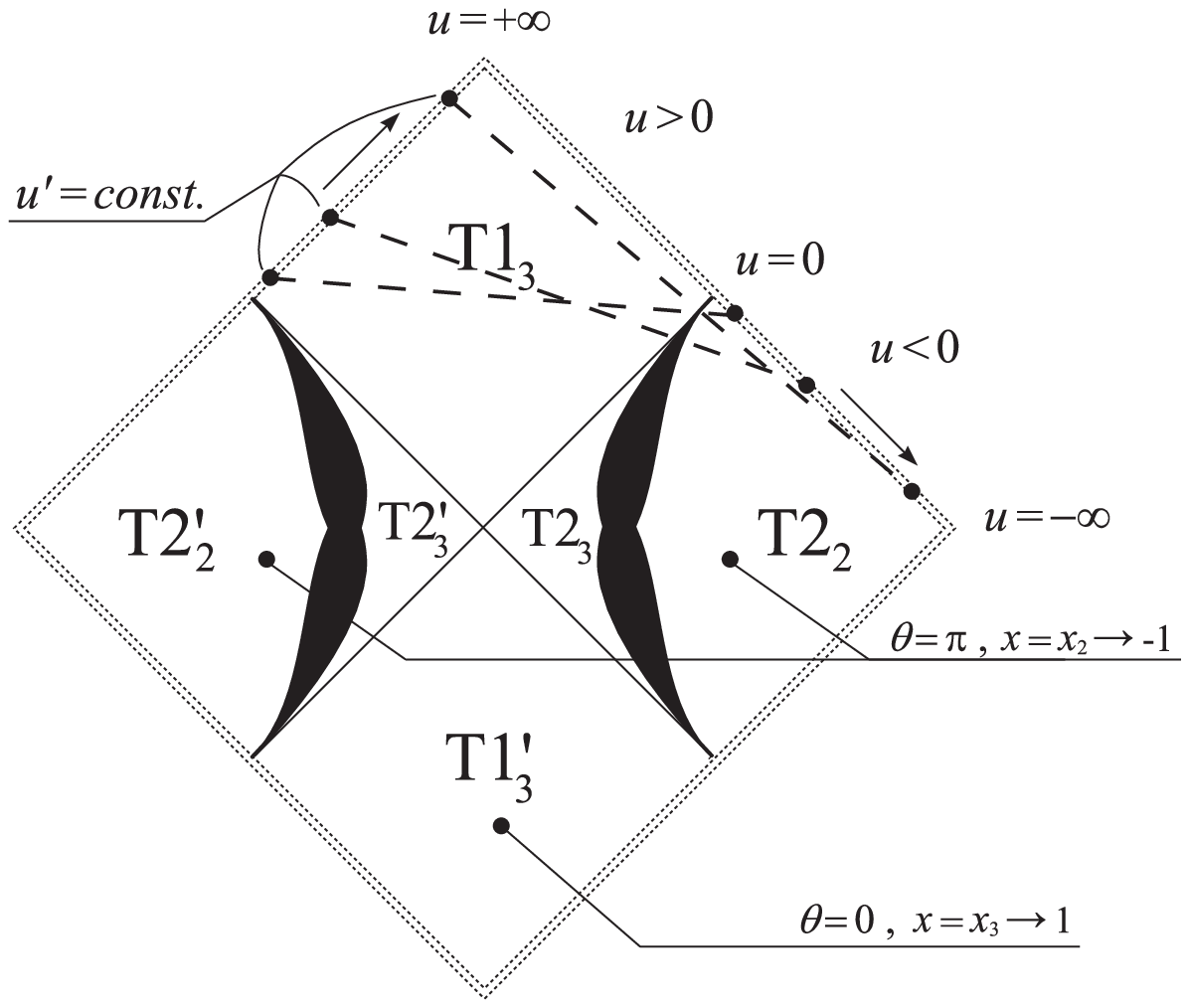}
    \caption{Conformal diagram representing the Schwarzschild limit $A \rightarrow 0$ and the supertranslation
     $\hat u' \rightarrow \hat u + \alpha$, from the perspective of the original Bondi time coordinate
    $\hat u$.\vskip 1.5em  }\label{FigSchwLim}\notag
  \end{center}
 \end{minipage}
 \begin{minipage}{0.47\linewidth}
  \begin{center}
   \includegraphics[height=7.5cm]{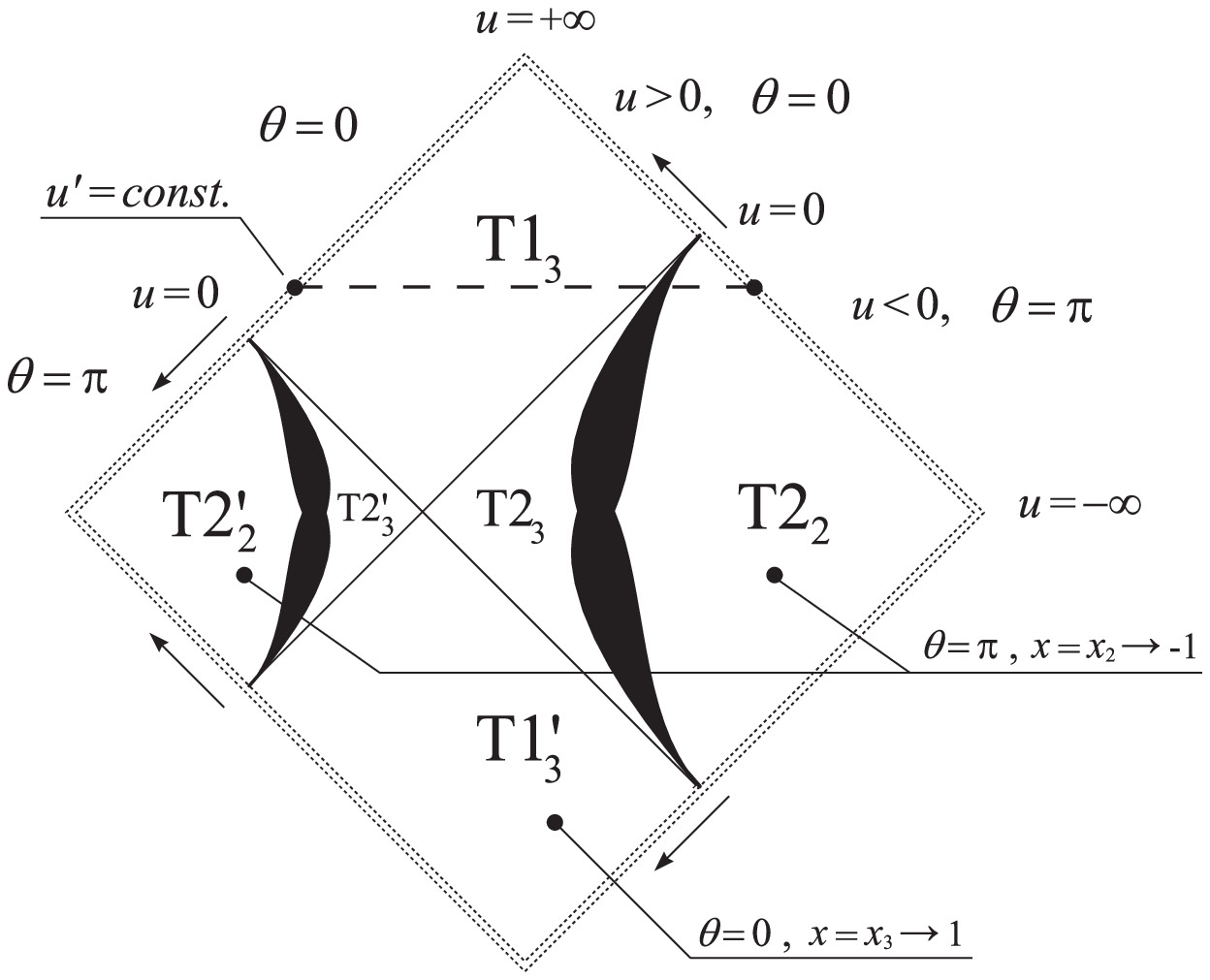}
   \caption{Same as figure \ref{FigSchwLim}, but from the perspective of the translated Bondi time coordinate
    $\hat u'$. The emerging conformal diagram of a single black hole (blocks $T2_3$, $T2_2$ and the black hole interior)
     is now clearly recognized.}\label{FigSchwLim2}\notag
  \end{center}
 \end{minipage}

\end{figure}

We start with the expansion of (\ref{Sdef}), using (\ref{mAexpansion-Gi}) and (\ref{mASconstExpansion}) :
\begin{align}\notag\label{SdefA}
\kappa_{ext}&=1 : \quad
 \quad S = -\arctanh x + \Big( u' - 2m'\ln(1{-}x) -\frac{m'}{1{-}x^2}   \, \Big) \,A + O(A^2) \\
\kappa_{in}&=1 : \quad
 \quad S = -\arctanh x + \Big( u' - 2m'\ln(1{+}x) -\frac{m'}{1{-}x^2}   \, \Big) \,A + O(A^2) \cmb
\end{align}
Inverting it, we obtain, with the help of (\ref{OmS}):
\begin{align}\notag
\kappa_{ext}&=1 : \quad
 \quad x(u',\Theta) = \cos\Theta + \big[\sin^2\!\Theta\,\big( u' {-} 2m' \ln(1{-}\cos\Theta) \,\big) -m' \big] \,A + O(A^2) \\\notag
\kappa_{in}&=1 : \quad
 \quad x(u',\Theta) = \cos\Theta + \big[\sin^2\!\Theta\,\big( u' {-} 2m' \ln(1{+}\cos\Theta) \,\big) -m' \big] \,A + O(A^2) \cmb
\end{align}
This expansion confirms the expected coincidence of $\Theta$ and $\theta$ in the limit $A \rightarrow 0$,
and therefore the poles $x=x_{2,3}$ correspond here to (see also figure \ref{FigSchwLim},\ref{FigSchwLim2}):
\begin{align}\label{poles-A}
\qquad \lim_{A \rightarrow 0} \Theta = \theta \quad \Longrightarrow    \quad
        \Big\{
     \begin{array}{lcl}
        \Theta{=}0 & \leftrightarrow &   x= \lim_{A\rightarrow 0} x_3 =1  \\[0.55em]
        \Theta{=}\pi & \leftrightarrow  & x= \lim_{A\rightarrow 0} x_2 = -1
    \end{array}
         \cmmb
\end{align}
Also, we can now express $\hat u$ (\ref{uhat}) as:
\begin{align}\notag
\kappa_{ext}&=1 : \quad
 \quad \hat u =  \frac{\cos \Theta}{A} + u' -2m' \cos\!\Theta - 2m' \ln(1{-}\cos\Theta) -3m' + O(A) \\ \notag
 \kappa_{in}&=1 : \quad
 \quad \hat u =  \frac{\cos \Theta}{A} + u' +2m' \cos\!\Theta - 2m' \ln(1{+}\cos\Theta) -3m' + O(A) \cmb
\end{align}
Apparently the limit of $\hat u$ for $A {\rightarrow} 0$ is diverging and is obviously not
the Bondi time for the Schwarzschild metric. However using a specific supertranslation, it can be corrected:
\begin{align}\label{superTr}\notag
\hat u \ \rightarrow \ \hat u' &= \hat u+\alpha \cma \\
& \alpha(\Theta,\phi) =  \ \Bigg\{
    \begin{array}{lcl}
        \kappa_{ext}&=1  \quad: &   \ds -\frac{\cos \Theta}{A} +2m' \cos\!\Theta + 2m' \ln(1{-}\cos\Theta) +3m'  \\[1em]
        \kappa_{in}&=1 \quad:   &  \ds -\frac{\cos \Theta}{A} -2m' \cos\!\Theta + 2m' \ln(1{+}\cos\Theta) +3m'
    \end{array}\cmma \qquad
\end{align}
which ensures that $\lim_{A \rightarrow 0} \hat u=u'$. The supertranslation is qualitatively depicted
on figure \ref{FigSchwLim} and \ref{FigSchwLim2}. The logarithmic term appears to be unavoidable and a divergence of this type is always
present regardless of higher order differences of the limiting scheme.
It seems to be a manifestation of presence of the conical singularity on the respective axial
segment; it always diverges on the pole(s) which correspond to the axial segment exhibiting a conical singularity
(recall (\ref{poles-A}) and figure \ref{FigSchwLim}).

Now when the limiting process is properly set up, we continue to examine the behaviour of  the news function and
the mass aspect. In general, we will use (\ref{SchwLimParamScale}), (\ref{SchwLimCoordScale}),
(\ref{SdefA}) and (\ref{superTr})
to obtain the $A \rightarrow 0$ asymptotic expansion of those quantities.

The $c_{,u}$ expression (\ref{c_u}) reduces to :
\begin{align}\notag
\kappa_{ext}&=1 :& \\\notag
\quad c_{,u}&= {-} \big(\! \cos\theta - 2\big) \cotan^2\!\frac{\theta}{2} \ m'A {}+{} \\ \notag
&\  +\cotan^2\!\frac{\theta}{2}\, \bigg(  2\cos\theta \, ( \cos\theta{-}2) \Big(\frac{u'}{m'} -2\ln (1{-}\cos\theta) \Big)
-\frac{3}{2}\cos^2\theta +\frac{7}{2} +2\,\frac{2{-}\cos\theta}{\sin^2\theta} \, \bigg)\, m'^2\!A^2 \,+\, O(A^3)
\\[.75em]
\kappa_{in} &=1 :& \\\notag
\quad c_{,u}&=-\big(\! \cos\theta+ 2 \big) \tan^2\!\frac{\theta}{2} \ m'A {}+{} \\ \notag
&\  +\tan^2\!\frac{\theta}{2}\, \bigg(  2\cos\theta \, ( \cos\theta{+} 2 ) \Big(\frac{u'}{m'} - 2\ln (1{+}\cos\theta) \Big)
-\frac{3}{2}\cos^2\theta +\frac{7}{2} +2\,\frac{2{+}\cos\theta}{\sin^2\theta} \, \bigg)\, m'^2\!A^2 \,+\, O(A^3)
\end{align}
where we have again expressed the result for the case of the conical singularity vanishing
either between ($\kappa_{in} {=}1$) or outside ($\kappa_{ext}{=}1$) of the accelerated black holes
, using (\ref{SchwLimParamScale}).
The behaviour near the poles is then:
\begin{align}\notag
\kappa_{ext}&=1: \quad \Bigg\{
\begin{array}{l}
\ds c_{,u}=\big(4m' \theta^{-2} + O(\theta^0) \big) A + \big(8m'^2 \theta^{-4} +O(\theta^{-2}) \big) A^2 + O(A^3) \\[0.55em]
\ds c_{,u}=\bigg( \frac{3}{2}\,m' (\pi{-}\theta)^2 + O\big((\pi{-}\theta)^4\big) \bigg)A -\bigg( \frac{3}{2}m'^2 + O\big((\pi{-}\theta)^2\big) \bigg) A^2  + O(A^3)
\cmma
\end{array} \\[1.8em]
\kappa_{in}&=1: \quad \Bigg\{
\begin{array}{l}\\[-2em]
\ds c_{,u}=\bigg( {-}\frac{3}{2}\,m' \theta^2 + O(\theta^4) \bigg)A -\bigg( \frac{3}{2}m'^2 + O(\theta^{2})\big) \bigg) A^2  +
O(A^3) \\[1.2em]
\ds c_{,u}=\Big({-}4m' (\pi{-}\theta)^{-2} + O\big((\pi{-}\theta)^0\big) \Big) A + \Big(8m'^2 (\pi{-}\theta)^{-4} +O\big((\pi{-}\theta)^{-2}\big) \Big) A^2 + O(A^3)
\end{array}
\end{align}
which is an expected result, consistent with an intuitive approach encouraged by
figure \ref{FigSchwLim}.  In both cases $c_{,u}$ is always regular at one pole regular
 and diverging at the other,
because the supertranslation cancels the symmetry of $\theta \leftrightarrow \pi{-}\theta$.  (By
regularity here we mean that  the integral $\int_{I_\epsilon} c_{,u}^2 \sin\! \theta \, d\theta $ exists
on some small neighbourhood $I_\epsilon$ of a given pole.)
As we already know,  $\theta{=}0 \leftrightarrow x{=}1$ and $\theta{=}\pi \leftrightarrow
x{=}-1$, and since in the C-metric we cannot eliminate
the conical singularity on both axes (axial slices), at least one pole must be singular.

We may now continue with the computation of the Bondi mass aspect $M$. First we check the
behaviour of the reduced mass aspect $\hat M$. Employing again the expansion (\ref{SdefA}) and the
supertranslation (\ref{superTr}) in equation (\ref{MH}), we find:
\begin{align}\label{MH-Schw}
\hat M =  m' + O(A) \cma
\end{align}
The limit of $\hat M$ is well behaved with the result being exactly
what one would expect from the limit (\ref{gSchwLim}) of the metric, compared to equation
(\ref{BondiMetric}).

However, as a more detailed computation shows, the terms
of $O(A)$ and higher diverge near both poles at least as $O(\theta^{-3})$ and $O\big( (\pi{-}\theta)^{-3}\big)$ respectively.
Therefore the integral of $\hat M$ over the $2-$sphere, i.e., the total mass of the system, is defined in the limit itself
only, while for an arbitrarily small
non-zero $A$, the integral does not exist.
This is not so surprising; we had a similar behaviour in the case of the news function $c_{,u}$.

Still, we might have expected that at
least near one pole, the situation could be made regular. This is not true
for $\hat M$, but as we show below, it can\footnote{Of course, the integral of $M$
 still does not, and cannot, exist.  If it existed, it would,
 according to equation (37) and the remark below it in \cite{tafel_2}, have to be equal to
the integral of $\hat M$, which is diverging.} be done in the case of $M$.

Proceeding to find $M$, we find :
\begin{align}\notag
\kappa_{ext}&=1 : \quad
\quad M=m'+ \bigg( 3\cos\theta \, \Big(1+2\ln(1{-}\cos\theta) -\frac{u'}{m'} \Big) +5 \bigg) \,  m'^2 A + O(A^2)
\\[.75em]
\kappa_{in}&=1 : \quad
\quad M=m'+\bigg( 3\cos\theta \, \Big(1+2\ln(1{+}\cos\theta) -\frac{u'}{m'} \Big) -5 \bigg) \,  m'^2 A + O(A^2)
\end{align}
This confirms that in the limit $A \rightarrow 0$ the Bondi mass equals the Schwarzschild mass.
Series expansion near the poles $\theta{=}0,\pi$ then reveals:
\begin{align}\notag
\kappa_{ext}&{=}1: \ \Bigg\{
\begin{array}{l}
\\[-1.2em]
\ds M=m'+ \bigg( \Big(  3+6\ln\!\frac{\theta^2}{2} -3\frac{u'}{m'} +5 \Big) \,  m'^2 + O(\theta^2) \bigg) A + O(A^2) \\[1em]\notag
\ds M=m'+\bigg( \Big(  -3-6\ln 2 + 3\frac{u'}{m'} +5 \Big) \,  m'^2  + O\big((\pi{-}\theta)^2\big) \bigg) A + O(A^2)
\cmma
\end{array} \\[1em]
\kappa_{in}&{=}1:  \ \Bigg\{
\begin{array}{l}
\ds M=m'+ \bigg( \Big(  3+6\ln2 -3\frac{u'}{m'} -5 \Big) \,  m'^2 + O(\theta^2) \bigg) A + O(A^2) \\[1em]
\ds M=m'+\bigg( \Big(  -3-6\ln\!\frac{(\pi{-}\theta)^2}{2} + 3\frac{u'}{m'} -5 \Big) \,  m'^2  + O\big((\pi{-}\theta)^2\big) \bigg) A +
O(A^2)\cmmb
\end{array}
 \end{align}
This seems rather surprising, since it appears as if the mass aspect $M$ was integrable over the whole $2-$sphere.
But again, a more detailed analysis reveals that the terms $O(A)^2$ and higher all diverge at the poles where $\kappa \neq 1$,
leading to the same qualitative behaviour as in the case of $c_{,u}$. The integral of $M$ over the entire $2-$sphere therefore does not
exist, as we have expected, unless we restrict to terms of $O(A)$ or to the limit $A{=}0$ itself.

An interesting question arises concerning the relation of the Schwarzschild limit $A \rightarrow 0$ and the
Minkowski limit $m \rightarrow 0$. We will shortly address this issue by investigating
the small mass limit of the coordinates used in the Schwarzschild limit.
First, we express the function $x$ as $x(w,\Theta)$, using the inverse of the expansion of
$Gi(x)$ (see (\ref{mAexpansion-Gi}). Assuming axial regularity, i.e. $\K=1{+}O(mA)$, this leads to:
  \begin{align}\label{xw-m}
       x(w,\Theta) &= \tnh(w{-}\K S) 
           =\frac{ \ch\!S \sh w  -  \sh\!S \ch w}{\ch\!  S \ch w -\sh\!  S \sh
           w}+O(mA)
             = \frac{ \cos\!\Theta \ch w + \sh w}{\ch w + \cos\!\Theta \sh w} + O(mA)\cmb
       \end{align}
Comparing this to the Schwarzschild limit formulas for $\theta$ and $u'$ coordinates (see (\ref{gSchwLimMetric}) and (\ref{wSchw}))
, we obtain the desired relation:
 \begin{align}\label{mS-rel}
\cos \theta =   \frac{ \cos\!\Theta \,\ch u'A + \sh u'A}{\ch u'A + \cos\!\Theta \,
\sh u'\!A} + O(mA) \cmb
   \end{align}
In terms of the complex stereographic coordinate $\xi$, related to the angle $\Theta$ via (\ref{stereog}), and
the coordinate $\zeta$ related in the same way to $\theta$, this
translates into a simpler formula:
 \begin{align}\label{mS-rel-stereog}
\zeta=\xi e^{-u'A} \cma
   \end{align}
in which we recognize the Lorentz boost along the $z$ axis, with                        
$\beta=-\tanh u'A$ being the velocity. This confirms the intuitive idea that the small
mass limit of Schwarzschild limit $A \rightarrow 0 $ and the Minkowski limit $m
\rightarrow 0$ are related by a Lorentz boost along the symmetry axis, with the
velocity increasing as $u'$ increases; the Schwarzschild limit corresponds to the
observer at rest with respect to the black hole -- particle, which, on the other hand is accelerating
along the axis in the Minkowski observer's coordinate frame. They should therefore be
related by a boost, with the velocity as a function of the acceleration and time.

Another way to understand this is to realize that different Bondi coordinates must be related
by a transformation belonging to the BMS group\footnote{See, for example \cite{tafel_2}, section 3.},
 which, roughly speaking, consists of boosts,
 rotations and supertranslations. Since in both limits the angular
coordinates were adapted to the symmetry axis, they must be related only by a pure
boost, or, in general, modulo some additional supertranslation.

\section{Bondi time limit $\bold u \rightarrow 0 $ and $\bold u \rightarrow \pm \infty $}
\label{ulimits}

The explicit formulas for the functions $\Gj(x)$ are given in the Appendix, equations
(\ref{gj-intexplicit-C2}) and (\ref{gj-intexplicit-C3}). They allow us not only to
numerically compute the mass aspect
and the total mass as their integral over the entire $\scri^+$, but also to obtain some analytic results.

Although the analytical computation of the total mass does not seem feasible in the general case, there exists a
well defined limiting behaviour for Bondi time $u \rightarrow 0^\pm$ and $u \rightarrow
\pm \infty$. These limits correspond to the observer located on $\scri^+$ approaching the event of the black hole
hitting $\scri$, or respectively, him moving to the time future or spatial infinity  (see
figure \ref{slices}). The total mass can be expressed using (\ref{bondiQ}), which in our case
leads to :
\begin{align}\label{BondiLimit}\notag
M(u) & =\frac{1}{2}\int^{\pi}_{0} \!  \hat M(u,\Theta) \sin\! \Theta\, d\Theta  = \int\limits_{
I}{\ds} J(u,x) \hat M(u,x)\, dx \quad , \
\begin{array}{l}
  I = \Big \{
  \begin{array}{l}
  u{>}0 : (x_u,x_3) \\[0.3em]
  u{<}0 : (x_2,x_u)
  \end{array}
  \\[1.2em]
  \quad \Gj(x_u)=uA\K
\end{array}
\\[0.6em]
 &J  = \sin \! \Theta\, \det \! \frac{\partial (u,\Theta)}{\partial (u,x)} = -\frac{\sin^2\Theta}{G^{\frac{3}{2}}\Gj \cos \Theta}
\, \mathop{=}_{(\alpha=0)} \,
 -\frac{u^2A^2\K^2}{\Gj^2G^{\frac{3}{2}}} \frac{\sign( \Gj \cos \! \Theta)}{\sqrt{\Gj^2{-}u^2A^2\K^2}}
\end{align}
where we have used $\hat M$ instead of $M$ to simplify the following calculations. These expressions allow us to
compute the various limits with respect to $u$ that we want.  We will have to use the first one for $u\rightarrow \pm \infty$
and the latter for the $u\rightarrow 0$ limit, in order to be able to swap $\int_I$ and
$\lim\limits_{u\rightarrow \pm\infty,0}$. Also the mass will be finite in the
first place in the $u{>}0$ or $u{<}0$ case only when we set the conicity parameter $\K$ to
have $\kappa_{in}{=}1$ or $\kappa_{ext}{=}1$ respectively. Then, we obtain the following limiting behaviour for the mass
$M(u)$:

\begin{align}\notag
\kappa_{in}=1 \ : \quad  M(u) \ = \ \frac{24 \, x_1^4 \, x_2^4 \, x_3^4}{35 A^8
 (x_3{-}x_1)^6(x_3{-}x_2)^6}\ \frac{1}{u^{7}} + O(u^{-8})
  \cmma \quad  \text{as } u \rightarrow +\infty  \\[1em]
\kappa_{ext}=1 \ : \quad  M(u) \ = \ \frac{24 \, x_1^4 \, x_2^4 \, x_3^4}{35 A^8
 (x_2{-}x_1)^6(x_3{-}x_2)^6}\ \frac{1}{u^{7}} + O(u^{-8})
  \cmma \quad  \text{as } u \rightarrow -\infty
\end{align}

For the limit of $u\rightarrow 0^\pm$ the evaluation of the integral is more complicated.
In order to obtain an explicit result, we restrict the supertranslation
freedom to the class of $\alpha = \frac{C\sin\Theta}{A\K}$. Since this is equivalent to $\Gj \rightarrow \Gj'=\Gj+C$, it allows us to use
the simplified formulas of the $\alpha{=}0$, which can be explicitly integrated, and then substitute $\Gj \rightarrow \Gj+C$.
Using this in the series expansion of $J\hat M$ in $u$ we obtain the following:
\begin{align}\notag\label{Mu0_limit}
 & M(u) \  = \ \int\limits_{I} J(u,x) M(u,x) dx  = \ \int\limits_{I}  J(u,x) \hat M(u,x) dx   \\[0.5em]\notag
 & \hskip 1.8em  =  \frac{1}{u} \!\! \int\limits_{\quad I_{|x_u=x_0}} \!\!\!\!\!  \Big(\  \frac{\, 3\K^2G'^3{-}24G'{-}6\K^2GG'G''{+}4\K^2G^2G'''}{96A^2G^2}
 - \frac{\Gj}{4A^2\K^2G^{\frac{3}{2}} } \ \, \Big) \, dx + O(1) \   \\[-3.5em]\notag
 & = \ \frac{1}{u}  \ . \
  \left\{ {\vbox{\hbox{\strut}\hbox{\strut}\hbox{}}} \right.
  \begin{array}{l} \ds  \\[2.9em]
   \text{ for } I=(x_2,x_u) \ \Leftrightarrow \ u{<}0  \ : \\[0.5em]
      \quad \ds -\frac{(x_1{+}x_3{-}x_2)\,\K_2}{2A^2(x_3{-}x_2)(x_2{-}x_1)} - \bigg [ \frac{24{-}3\,\K_{2}^2 G'^2}{96A^2G}+\frac{G'''x\K_{2}^2}{24A^2} -
              \frac{C_2^2}{8A^2\K_{2}^2} \bigg ]_{|x=x_0}
                  \\[3em]
   \text{ for } I=(x_u,x_3) \  \Leftrightarrow \ u{>}0 \ : \\[0.5em]
      \quad \ds -\frac{(x_1{+}x_2{-}x_3)\,\K_3}{2A^2(x_3{-}x_2)(x_3{-}x_1)} - \bigg [ \frac{24{-}3\,\K_{3}^2 G'^2}{96A^2G}+\frac{G'''x\K_{3}^2}{24A^2} -
              \frac{C_3^2}{8A^2\K_{3}^2} \bigg ]_{|x=x_0} \\[0.3em]
              \\[2.3em]
  \end{array}
  \left. {\vbox{\hbox{\strut}\hbox{\strut}\hbox{}}} \right\} \quad + O(1) \cma
  \\[-5.5em]\\\notag 
\end{align}
where $x_0$ is chosen so that $\Gj(x_0){=}0$, and the quantities $C_{i}=0$
are the constant terms in the expansion of $\Gj$, see (\ref{gj-intexplicit-C2}), (\ref{gj-intexplicit-C3}), (\ref{Gj_i}), while
$\K_{i}$ is a conicity parameter corresponding to the physical conicity $\kappa_i{=}1$ (see
(\ref{CMcon})). Again, the conicity has been set for the conical singularity to disappear on the respective
axis segments (either $x=x_2$ or $x=x_3$) in order for the result to be finite in the first place.

To summarize this, we have shown that in the absence of a conical singularity, the Bondi mass behaves as
$1/u^7$ for large null time, $u \rightarrow \pm \infty$, and as\footnote{At least for the $\alpha=\frac{C \sin \Theta}{A\K}$ class
of supertranslations.} $1/u$ for time close to the event of the black
hole reaching $\scri^+$ when $u \rightarrow \pm 0$.

The leading coefficients in the above asymptotic expansions are strictly positive, causing
the Bondi mass to be strictly non-increasing, as required by general theorems,
demonstrated by, e.g. (\ref{bondiQ}). In the $u \rightarrow \pm \infty$ case this is
obvious; for the $u \rightarrow 0^\pm$ see the proof in \ref{A-bondimass}.

\section{Regularity near the poles in the general case}
\label{genpolarlimit}

In sections \ref{SmallMassLim} and \ref{SchwLim}, concerning the small mass and small acceleration limit,
we have encountered a rather similar behaviour of the Bondi mass and the news function with respect to the
regularity of the corresponding segment of the symmetry axis; the above mentioned asymptotic quantities were
only diverging at the pole if and only if the respective axial segment contained a conical singularity.
This suggests that such behaviour may be preserved even in the general case, and we will investigate it in this section.

 To analyze the Bondi mass (\ref{bondiM-CM}) and the news function (\ref{n_c_rel}) directly is rather involved, in the
sense that an analytical expression of $M$ and $c_{,u}$ in terms of the Bondi coordinates
$\hat u,\Theta$ is extremely complicated\footnote{This is so because we would have to express
$x$ as a function of $\Theta$ and $\hat u$, using the inverse of $\Gj(x)$, and then
substitute it into $G$ and its derivatives.}. However, as we show below, the situation
near the poles $x_{2,3}$ can be investigated analytically, even in the general case.

To obtain the behaviour of the news function $c_{,u}$ and of the Bondi mass $M$ near
the poles\footnote{See (\ref{poles-m}), (\ref{poles-A} for $x_i$ correspondence.} at $\Theta{ = }0$ and $\pi$,
 in the full relativistic case, we substitute the expansions for $G(x)$,
 considering the case that $G(x)$ is a third order polynomial:
\begin{align}\notag
    G(x)&=G(x_2{+}\xi_2) \quad\quad \rightarrow &
          G(x) &= \hphantom{-} G'(x_2)\,\xi_2+ G''(x_2) \frac{\xi_2^2}{2} + G'''(x_2)
        \frac{\xi_2^3}{6}
     \\[0.5em]
    G(x)&=G(x_3{-}\xi_3) \quad\quad \rightarrow &
              G(x) &= {-}G'(x_3)\,\xi_3+ G''(x_3) \frac{\xi_3^2}{2} - G'''(x_3)
        \frac{\xi_3^3}{6}
\end{align}
and analogically for its derivatives, into (\ref{c_u}) and (\ref{bondiM-CM}) respectively. The conicity parameter is expressed in terms of the physical
conicity via (\ref{CMcon}). For $c_{,u}$, this leads to:
\begin{align}\label{cu-polar}
c_{,u}=\frac{1}{2\sin^2 \! \Theta}\, \big( 1{-}\kappa^2_{i } \big) \, +\,
\frac{G'''_{i}\,\kappa^2_i}{G'_i\sin^2\!\Theta} \,\xi_i^2 \, +\,  \frac{1}{\sin^2\!\Theta}\ O(\xi_i^3)
\cmma \quad i=2,3 \cma \quad G^{(j)}_i \equiv G^{(j)}(x_i) \cmmb
\end{align}
Noting that:
\begin{align}\label{sinexpansion}
\frac{1}{\sin\Theta} = \frac{1}{A\K (\hat u{-}\alpha)} \bigg( \frac{1}{\sqrt{\xi_{2,3}}} +
O\big(\sqrt{\xi_{2,3}}\big) \bigg) \cmma
\end{align}
we see that only\footnote{Assuming  that $\lim_{\Theta \rightarrow (0,\pi) }\alpha  \neq u$, which is where the black
holes approach the $\scri^+$. }
 the first term in (\ref{cu-polar}) can diverge for $\kappa_{2,3} \neq 1$.
In other words this means
 that on the axial slices $x{=}x_{2,3}$ we indeed can again, by choosing the axis to be regular either
at $x{=}x_2$ or at $x{=}x_3$, eliminate the singular behaviour of $c_{,u}$ on
$\scri^+$, either for $\hat u{-} \alpha >0$, or $\hat u{-} \alpha<0$ respectively\footnote{See also
the conformal diagrams on figure \ref{x3slice},\ref{x2slice},\ref{slices}.}.

Now for the Bondi mass aspect $M$, we again use the above mentioned approach and arrive to:
\begin{align}\label{M-polar}\notag
M =& \frac{1}{4} \frac{\epsilon_i \kappa_i (\kappa_i^2{-}1)}{A \sqrt{|G'_i|} \, \sin^3 \Theta} \
 \frac{1}{\sqrt{\xi_i}}  \ +\ \frac{1}{4} \frac{\kappa_i^2{-}1}{\sin^4\Theta} \Big( \sin
 \Theta \cos \Theta \, \alpha' -\sin^2 \Theta \, \alpha'' +u{-}\alpha \Big) \  \\\notag
&+ \ \frac{3}{16} \, \frac{\kappa_i(\kappa_i^2{-}1)}{A \sin^3\! \Theta} \,
\frac{G_i''}{|G_i'|^{\frac{3}{2}}}\ \sqrt{\xi_i} \ + \  \frac{\epsilon_i \kappa_i}{384\,A}
\,
\frac{G'_i G'''_i \, (72\kappa_i^2{-}40) - 15 \, G_i''^2
 (\kappa^2_i{-}1)}{|G_i'|^\frac{5}{2} \sin^3\!  \Theta} \ \xi_i^{\frac{3}{2}} \  \\\notag
&\hskip 10em - \ \frac{1}{4} \frac{\kappa_i^2 G'''_i}{G_i' \sin^4\!\Theta} \, \big(\sin\!\Theta \cos\!\Theta
\, \alpha' - \sin^2\!\Theta \, \alpha'' + u{-}\alpha \big ) \ \xi_i^2 \  \ \\
 & \quad - \ \bigg[ \frac{AG_i'''}{\sqrt{|G'_i|} \sin^5\!\Theta} \big( \sin\!\Theta \, \alpha' {+}
 \cos\!\Theta\, (u{-}\alpha) \big)^2 + \frac{1}{48A} \frac{G_i'''
 G''_2}{|G_i'|^{\frac{5}{2}} \sin^3\Theta} \bigg] \, \xi_i^{\frac{5}{2}} \ + \ O(\xi_i^3)
 \cmma
\end{align}
where $\epsilon_i= +1,-1$ for $i=2,3$. Noting (\ref{sinexpansion}) we see that also in
this case, for $\kappa_{2,3}=1$, the mass aspect $M$ does not diverge near the poles.

This in fact even holds for any $\alpha(\Theta)$ which is at least\footnote{So its second derivative in
(\ref{M-polar})
  is defined.} $\mathcal{C}^2$ on a neighbourhood of
$\Theta{=}0$ and $\Theta{=}\pi$, as can be seen by analyzing the $\Theta$ dependent terms
in the numerator:
\begin{align}\label{AB_reex}
A(\Theta) \equiv \sin
 \Theta \cos \Theta \, \alpha' -\sin^2 \Theta \, \alpha'' +u{-}\alpha \cmma \quad
B(\Theta) \equiv  \sin\!\Theta \, \alpha' {+}
 \cos\!\Theta\, (u{-}\alpha) \cmmb
\end{align}
For the situation near $\Theta=0$ we :

\begin{enumerate}
 \item[{I)}] First assume that $\alpha$ is bounded at $\Theta = 0$, and rewrite
 $A$ and $B$ as:
\begin{align}
A(\Theta)= 5(\alpha\sin\Theta)'\cos\Theta + \alpha\sin^2\Theta - (\alpha\sin^2\Theta)''
 +u-4\alpha   \cmma
 \quad B(\Theta)=
(\alpha\sin\Theta)' + (u-2\alpha)\cos\Theta
\end{align}
 Now, there
is a particulary nice limiting property of the expressions $(\alpha\sin\Theta)'$  and
$(\alpha\sin^2\Theta)''$ at $\Theta_0{=}0$ :
\begin{align}\notag\label{sa_limit}
 & \big( \sin\!\Theta \, \alpha(\Theta) \big)'_{|\Theta=\Theta_0} = \lim_{\epsilon
\rightarrow 0} \frac{ \sin (\Theta_0{+}\epsilon)\,  \alpha(\Theta_0{+}\epsilon) -  \sin
(\Theta_0) \, \alpha(\Theta_0)}{\epsilon} \ \ \underset{\Theta_0=0}{=} \ \
 \lim_{\epsilon
\rightarrow 0}\frac{ \sin (\epsilon)\,  \alpha(\epsilon) }{\epsilon} = \alpha(0)
\\
& \big( \sin^2\!\Theta \, \alpha(\Theta) \big)''_{|\Theta=\Theta_0=0} =
 \lim_{\epsilon\rightarrow 0} \frac{\sin^22\epsilon \ \alpha(2\epsilon) -2\sin^2\!\epsilon \
 \alpha(\epsilon) -\sin^2 0\ \alpha(0)}{\epsilon^2} = 2\alpha(0) \cmma
\end{align}
which, together with (\ref{AB_reex}) ensures the finiteness of $A$ and $B$ at
$\Theta{=}0$.

\item[II)] In the alternative case, of diverging $\alpha$, $\lim_{\Theta \rightarrow 0}
\alpha(\Theta)= \pm \infty$, and we can rewrite $A$ and $B$ in terms of $\beta{=}1/\alpha$  :
\begin{align}
A(\Theta)= \frac{( \beta\sin^2\Theta)''}{\beta^2} - 2\frac{(\beta\sin\Theta)'^2}{\beta^3}
-\frac{(\beta\sin\Theta)'}{\beta^2}\cos\Theta + \frac{\sin^2\Theta}{\beta} + u
  \cmma \quad
B(\Theta)= -\frac{(\beta\sin\Theta)'}{\beta^2} + u\cos\Theta \cmmb
\end{align}
Using the same idea as in the previous case, we see from (\ref{sa_limit}) that both $A$ and
$B$ diverge at worst as $1/\beta$. But this does not spoil the finiteness of
(\ref{M-polar}), since $A$ and $B$ only occurs as $ \frac{ A(\Theta)}{\sin^4\!\Theta} $
and \mbox{$ \frac{ B^2(\Theta)}{\sin^5\!\Theta} $}  (see (\ref{bondiM-CM})), and according
to (\ref{sinexpansion}), these $1/\sin^n(\Theta)$ factors are  more than enough to
compensate for the
\smallskip divergence of the excessive $\alpha$ in the numerator.
\end{enumerate}
The case of $\Theta \rightarrow \pi$ is completely analogous and leads to the same
conclusion.

\begin{figure}[h]
  \begin{center}
\subfigure[$\kappa_{in}{=}1$]{
   \includegraphics[height=4cm]{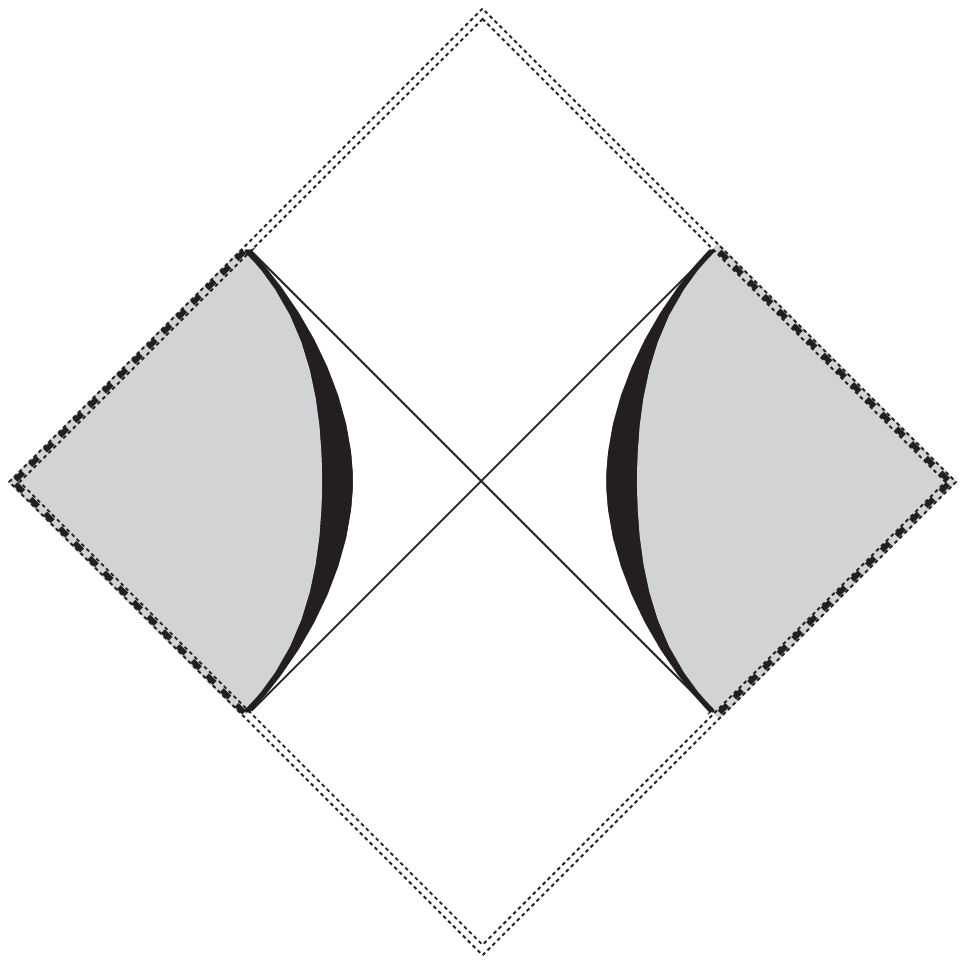}
   \label{confschem1}
}\hskip2em
\subfigure[$\kappa_{ext}{=}1$]{
   \includegraphics[height=4cm]{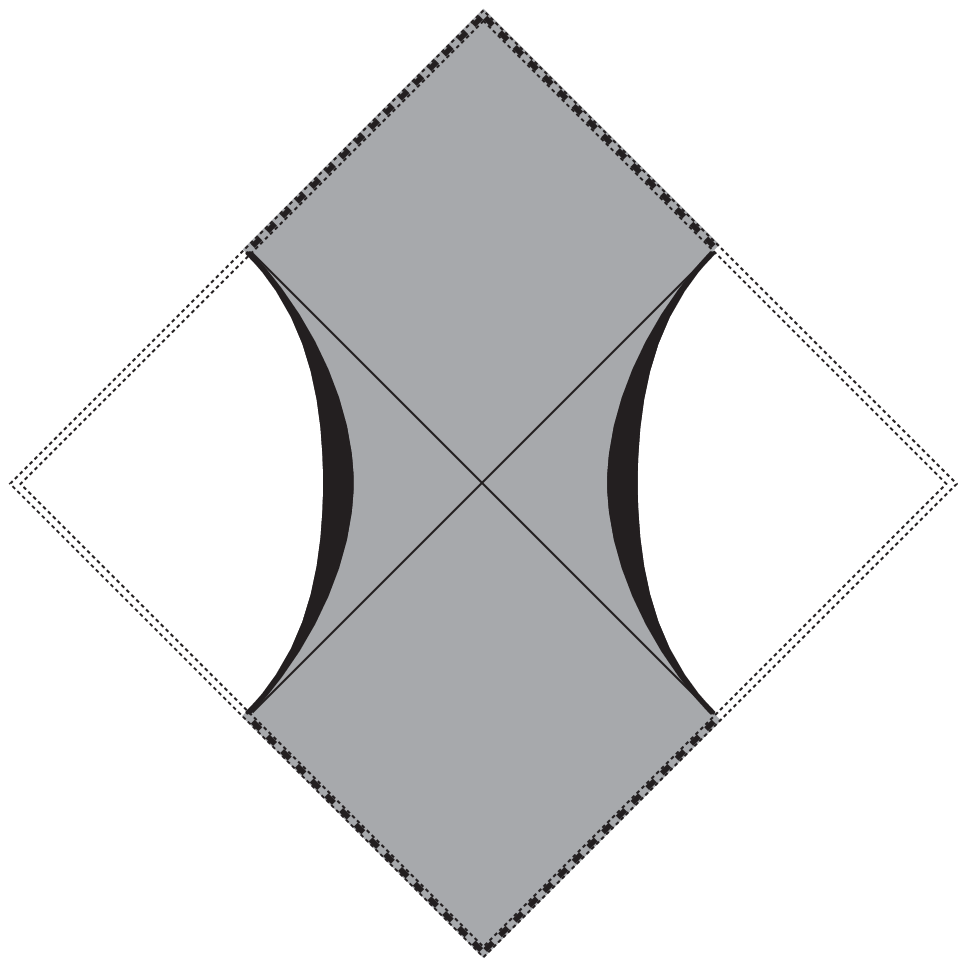}
   \label{confschem2}
}\hskip2em
\subfigure[$\kappa_{ext}{ \neq 1}$, $\kappa_{in}{ \neq 1}$, \newline note that $\kappa_{in}{>}\kappa_{ext}$] {
    \includegraphics[height=4cm]{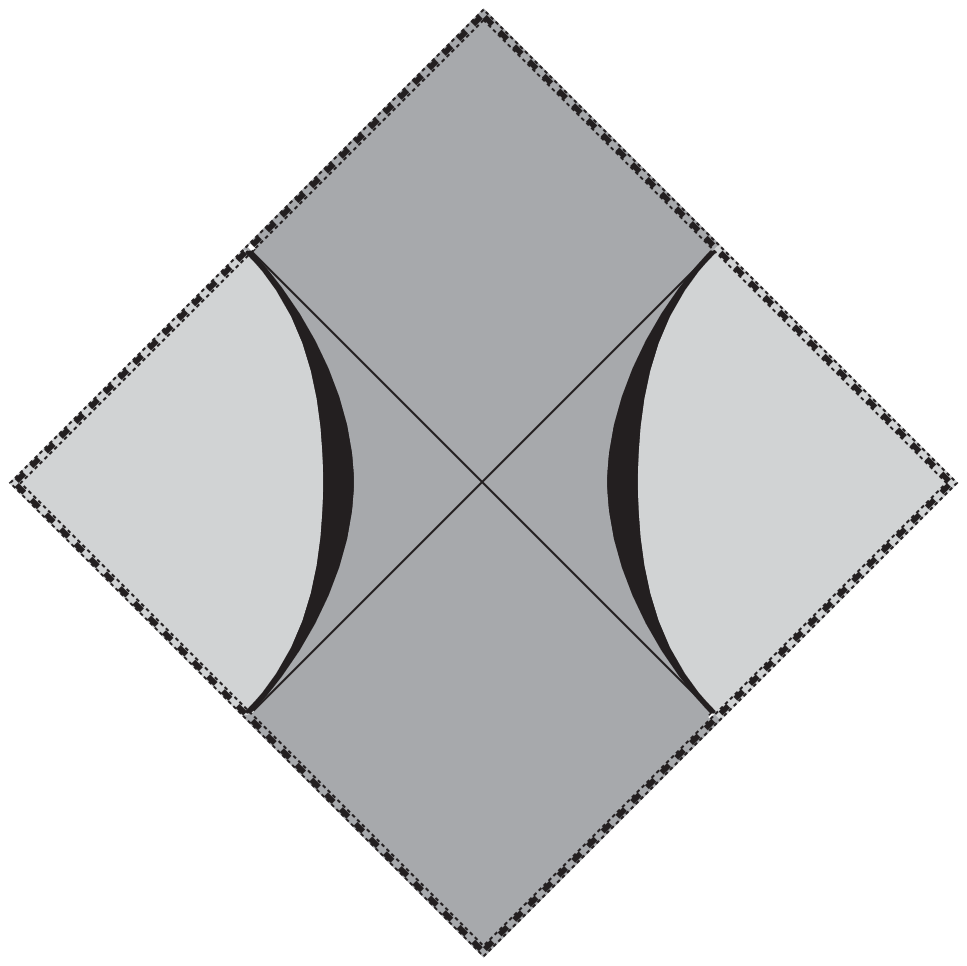}
    \label{confschem3}
}
\caption{Schematic conformal diagrams for the three possible cases of the conical singularity
location
and it's influence on the $\scri^+$ regularity. See also conformal diagrams of four prototypes of a general
boost-rotation symmetric spacetime given in figures 3-6 in \cite{bicak_2}.}\label{confschem}
\end{center}
\end{figure}

Therefore, as illustrated in figure \ref{confschem}, even in the general C-metric case, the conclusions of sections \ref{SmallMassLim} and \ref{SchwLim}
hold; the Bondi mass and the news function cannot be made regular if the corresponding axial segment
contains a conical singularity, and conversely, if the axial segment is regular, those quantities are regular
and integrable at the corresponding pole.

\ack
We would like to thank J Bi\v c\'ak for helpful comments and suggestions.
P.S. is also grateful for stimulating experiences at the University of New Mexico, Albuquerque, NM, U.S.A.
and the Albert Einstein Institute, Golm, Germany, where this work was partly done,
and also wishes to thank his dissertation advisor P Krtou\v s for guidance and discussion.



\appendix
\section{Some details of the computations}
\label{A-cmp}

To obtain the Bondi mass aspect, the news function and to perform the asymptotic calculations on $\scri^+$ in general,
the partial derivatives corresponding to the coordinate transformation
$(\hat u, \Theta) \leftrightarrow (w,x)$ are often needed.
Here, according to equations (\ref{uhat}) and (\ref{Sdef}), the
complete transformation Jacobian for the general case $ \alpha {\equiv} \alpha(\Theta,\phi) {\neq } 0\,$ is given:
\begin{align}\label{parcM}\notag
J=\left( \begin{array} {cc}
    \ds \Diff{\hat u}{w} & \ds\Diff{\hat u}{x} \\[1.2em]
    \ds    \Diff{\Theta}{w} & \ds \Diff{\Theta}{x}
\end{array} \right)
=& \ \frac{\sin \Theta}{\K G}  \left( \begin{array} {ccc}
    \ds  \frac{\cos\Theta}{A\K}\,G \Gj + G\Diff{\alpha}{\Theta} &\!\!,\!&
    \ds  \frac{1}{A\sqrt{G}} - \frac{\cos\Theta}{A\K}\, \,\Gj - \Diff{\alpha}{\Theta}
    \\[1.2em]
    \ds G &\!\!\!& \ds -1
\end{array} \right)
\\[2em]
J^{-1}= \left( \begin{array} {cc}
    \ds \Diff{w}{\hat u} & \ds\Diff{w}{\Theta} \\[1.2em]\notag
    \ds    \Diff{x}{\hat u} & \ds \Diff{x}{\Theta}
\end{array} \right)
=& \ \frac{\K }{\sin \Theta}  \left( \begin{array} {ccc}
    \ds  A \sqrt{G} &\!\!,\!&
    \ds  1- \frac{\cos \Theta} {\K}\,\sqrt{G} \Gj - A \sqrt{G} \, \Diff{\alpha}{\Theta}
    \\[1.2em]
    \ds A G^{\frac{3}{2}} &\!\!,\!& \ds -\frac{ \cos\Theta  }{\K} \,G^{\frac{3}{2}} \Gj  -
    AG^{\frac{3}{2}} \, \Diff{\alpha}{\Theta}
\end{array} \right)
\\[1em]
\det J =& \ -\frac{\sin^2\Theta}{A\K^2 G^{\frac{3}{2}}}
\end{align}
In sections \ref{SmallMassLim}, \ref{SchwLim} on limits, we used the $mA$ series of various expressions used in the general
calculations. For reference, the most important formulas are included here:
\begin{align}\label{mAexpansion-Gi}
Gi&=\arctanh x + \Big( \frac{1}{1{-}x^2}+\ln(1{-}x^2)  \Big ) \, mA +
  \bigg({-}\frac{x\,(8x^4{-}25x^2{+}15)}{2 (1{-}x^2)^2} +\frac{15}{4} \ln \frac{1{-}x}{1{+}x}
   \bigg)m^2\!A^2  \\[0.5em]\notag
   & \hskip25em + O\big(m^3\!A^3\big)         \\[0.75em]\label{mAexpansion-Gj}
\Gj&=\frac{x}{\sqrt{1-x^2}} + \frac{3x^2\!{-}2}{(1{-}x^2)^{\frac{3}{2}}}\, mA +
    \bigg( \frac{x\,(23x^4{-}35x^2{+}15)}{2(1{-}x^2)^{\frac{5}{2}}} - \frac{15}{2}\arcsin x \bigg) m^2\!A^2  {}
    \\[0.5em]\notag
                    & \hskip15em  - \frac{35x^8{-}280x^6{+}560x^4{-}448x^2{+}128}{2(1{-}x^2)^{\frac{7}{2}}}\,m^3\!A^3  +
                     O\big(m^4\!A^4\big)
\end{align}
\begin{align}\label{mAexpansion-x}
x&=\frac{\Gj}{\sqrt{1{+}\Gj^2}} + \frac{2{-}\Gj^2}{1{+}\Gj^2}\, mA +
      \bigg( 15 \arcsin \! {\txts \frac{\Gj}{\sqrt{1{+}\Gj^2}}}+ 5\Gj^3 - 27\Gj \bigg) \frac{m^2\!A^2}{2 (1{+}\Gj^2)^{\frac{3}{2}}}  +    O\big(m^2\!A^2\big)
        \\[1em]\notag
\end{align}
\begin{align}\label{kappaExp}\notag
     \kappa_{ext}&=\K (1-2mA)+ O(m^2\!A^2) \cma \qquad
     \kappa_{in}=\K (1+2mA)+ O(m^2\!A^2) \\[0.3em]
 & \longleftrightarrow \ \K=\kappa_{in} \big(1-2mA + \frac{15}{2}m^2\!A^2 -32m^3\!A^3 \big)+ O(m^4\!A^4)  \\\notag
 & \qquad \qquad \ = \kappa_{ext} \big(1+2mA +\frac{15}{2}m^2\!A^2 +32m^3\!A^3 \big)+ O(m^4\!A^4)
   \cma
\end{align}
where $x_i$ are the roots of the $G(x)$, $\K$ is the conicity parameter and $\kappa_{in}$,
$\kappa_{ext}$ is the conicity of the corresponding axial segment, see also \ref{A-conicity}.
Interestingly, the conicity parameter $\K$ can, in the light of (\ref{kappaExp}), be interpreted as an average
of the external and internal physical conicity, up to the second order of $A$.

\section{Conicity of the C-metric}
\label{A-conicity}

For a general axially symmetric $2-$space, $g_2=g_{RR}\,dR^2+g_{\phi\phi}\,d\phi^2$,
 with the coordinate $R$ such that the axis is located at $R{=}0$, the definition of
the conicity leads to the formula :
\begin{align}\label{congen}
\kappa=\ \lim_{\text{`distance to axis'}\rightarrow 0}\ \frac{\text{`circumference'}}
{2\pi{\times}\text{'distance to axis'}} = \lim_{R \rightarrow 0}
{\frac{{ \frac{\partial}{\partial R} }\sqrt{g_{\phi\phi}} }{ \sqrt{g_{RR}}}}_{\Big|{R=0}}\cmb
\end{align}

In the case of the C-metric (\ref{CM}), we find that:
\begin{align}\label{CMcon}
\kappa_{1,2,3}=\frac{\K}{2} |G'|_{|{x=x_{1,2,3}}} \cma
\end{align}
so the $\K$ parameter can really adjust $\kappa$ on a specific segment of the symmetry
axis. For the domain discussed here\footnote{Segment $x=x_1$ does
not lie in our spacetime; see beginning of section \ref{Cmetric-cf} and also figure (\ref{CMsquare}).},
the axis segment between the
particles lies at $x=x_3$ and the segment outside lies at $x=x_2$. We will therefore use
$\kappa_{ext} \equiv \kappa_2$ and $\kappa_{in} \equiv \kappa_3$ as synonyms in the text.

\section{Horizon Area}

The Schwarzschild limit $A \rightarrow 0$ was done holding the horizon area constant. Using a straightforward
integration, we find:
\begin{align}
  {\mathcal A}= \int\limits_{{\partial}S_2} \! \! \sqrt{g} \, d S_2 = \int^{2\pi}_{0}\!\! \int^{x_3}_{x_2} \!\! \sqrt{g_{xx} g_{\phi\phi} } \ dx d\phi =
  \frac{2\pi\K}{A^2} \frac{x_3{-}x_2}{(x_2{-}x_1)(x_3{-}x_1)} \cmb
\end{align}
Performing an expansion in $A$ leads to:
\begin{align}\label{areaExpansion}\notag
\kappa_{ext}&=1 : \quad
\quad M^2_0 \equiv \frac{{\mathcal A}}{16\pi}=m^2+ 2m^3A+26m^4 A^2 + 69m^5 A^3 + O(A^4)
\\[.75em]
\kappa_{in}&=1 : \quad
\quad M^2_0 \equiv \frac{{\mathcal A}}{16\pi}=m^2- 2m^3A+26m^4 A^2 - 69m^5 A^3 + O(A^4)
\end{align}
where $M_0$ is the mass of a Schwarzschild black hole with the same horizon area ${\mathcal A}$ as the accelerated black hole of the C-metric.
Inverting this series, we finally obtain (\ref{SchwLimParamScale}) :
\begin{align}\notag\label{mASconstExpansion}
\kappa_{ext}&=1 : \quad
\quad m=m' - m'^2 A -\frac{21}{2} m'^3 A^2 + \frac{71}{2} m'^4 A^3 + O(A^4)
\\[.75em]
\kappa_{in}&=1 : \quad
\quad m=m' + m'^2 A -\frac{21}{2} m'^3 A^2 - \frac{71}{2} m'^4 A^3 + O(A^4)
\end{align}

\section{Explicit formula for the $\Gj $ function}
\label{A-Gj}

\begin{figure}[h]
\begin{center}
\includegraphics[height=9cm]{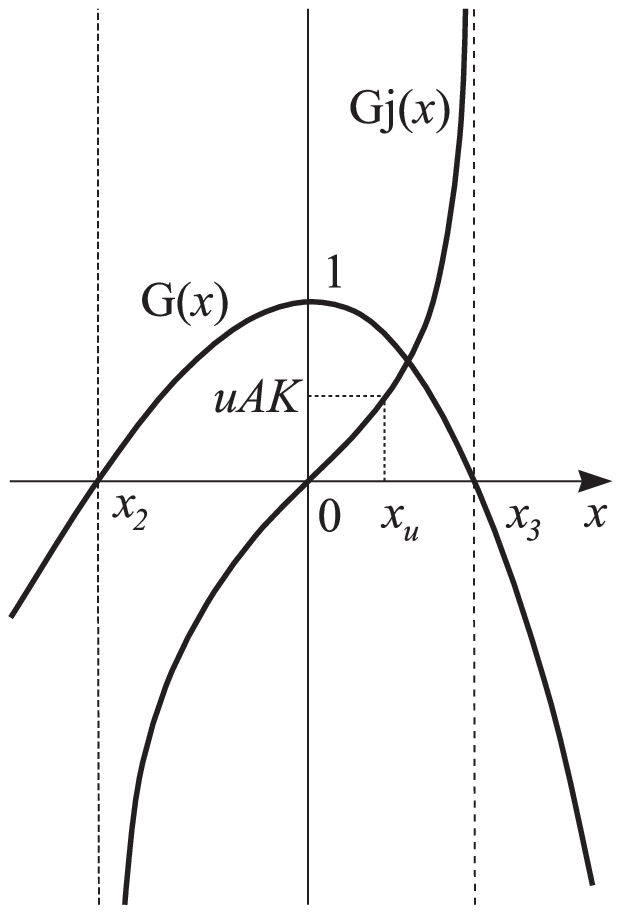}
\caption{Plot of the $G(x)$ and $\Gj(x)$ functions.  } \label{FigGGj}\notag
\end{center}
\end{figure}

The function $\Gj(x)$ is a crucial part of the formula for the Bondi time $\hat u$ and propagates
into other results as well.
While there is no problem with its qualitative description (see figure (\ref{FigGGj})) or numerical computation,
an analytical expression would certainly be helpful as well.
Fortunately, it appears to be possible to express the defining integral
\begin{align}\notag
\Gj(x) & = \int \! \frac{dx}{G(x)^{\frac{3}{2}}} = \frac{1}{a^{\frac{3}{2}}} \int
 \frac{dx}{\big ( (x{-}x_1)(x{-}x_2)(x_3{-}x) \big )^{\frac{3}{2}}} \equiv
 \frac{1}{a^{\frac{3}{2}}} \int
 \frac{dx}{( \xi_1\xi_2\zeta_3 )^{\frac{3}{2}}}
\end{align}
in an explicit form using elliptic functions. Here, we present two equivalent forms\footnote{
The third obvious one related to $x_1$ could be obtained as the last cyclic permutation of the indices of $x_i$.},
differing only by an integration constant:

 \begin{align}\label{gj-intexplicit-C2}\notag
\Gj_{(2)}(x)= \, & {-}b\,  d^{\frac{1}{2}}_{21}\, (d_{21}^2{+}d_{32}^2{+}d_{31}^2) \, \E\!\big(z_2,k_2 \big)\
  + \
 b\, d^{\frac{1}{2}}_{21}(d_{31}{+}d_{21}) \, \F\! \big(z_2,k_2\big)
\\[0.5em]\notag
& \hskip5em + \ b\, \frac{ \xi_1 \xi_2 d_{21}^2 + \xi_2 \xi_3 d_{32}^2 + \xi_3 \xi_1 d_{31}^2}
 {\sqrt{\xi_1 \xi_2 \zeta_3} } \ + \ C_2 \\[0.4em]
& \hskip2em \text{with   } z_2=\sqrt{\frac{\xi_2}{d_{32}}} = \sqrt{\frac{x-x_2}{d_{32}}} \cma \  k_2= i\, \sqrt{\frac{d_{32}}{d_{21}}}
 \cmma
 \end{align}
\\
\begin{align}\label{gj-intexplicit-C3}\notag
\Gj_{(3)}(x)= & \hphantom{-}b \, d_{32}^{\frac{1}{2}} (d_{32}^2{+}d_{31}^2{+}d_{21}^2)\, \E \!\big(z_3,k_3\big)\
 + \
 b \,  d_{32}^{\frac{1}{2}} (d_{32}{-}d_{21}) \,\F \!\big(z_3, k_3\big)
\\[0.5em]\notag
& \hskip5em + \ b\, \frac{ \xi_1 \xi_2 d_{21}^2 + \xi_2 \xi_3 d_{32}^2 + \xi_3 \xi_1 d_{31}^2}
 {\sqrt{\xi_1 \xi_2 \zeta_3}}  \ + \ C_3  \\[0.4em]
& \hskip2em  \text{with   } z_3=\sqrt{\frac{\zeta_3}{d_{31}}}
= \sqrt{\frac{x_3-x}{d_{32}}} \cma \  k_3= \sqrt{\frac{d_{31}}{d_{32}}}
  \cmma
\end{align}
\\
\begin{align}\notag
\text{and we have used the abbreviations }\quad \frac{2}{b}  = & a^{\frac{3}{2}} d_{21}^2 d_{31}^2 d_{32}^2 \cmma \\\notag
 \xi_i=x{-}x_i \cma \quad  \zeta_i={-}\xi_i & \cma \quad  d_{ij}=x_i{-}x_j
\cmma
\end{align}
\\
also with $\F(z,k)$ and $\E(z,k)$ being the incomplete elliptic integrals of the first and second
kind respectively\footnote{$F(z,k)= \int_0^z \frac{dt}{\sqrt{1-t^2}\sqrt{1-k^2 t^2}}$, $E(z,k)=
\int_0^z \frac{\sqrt{1-k^2 t^2}}{\sqrt{1-t^2}} \scs dt$ .}.
The two different forms are useful for series expansion as $x{\rightarrow} x_2$ from right and $x{ \rightarrow }x_3$ from left,
 with the advantage of $\xi_1$, $\xi_2$, $\zeta_3$ being always positive.
In the case of the gauge (\ref{GPar-b}) we also have additional relations:
\begin{align}\label{xi-indents}
2mA=a \cma \quad x_1+x_2+x_3=-\frac{1}{a} \cma \quad x_1x_2+x_1x_3+x_2x_3 = 0 \cma \quad
x_1 x_2 x_3 = \frac{1}{a} \cmmb
\end{align}
The expressions (\ref{gj-intexplicit-C2}), (\ref{gj-intexplicit-C3}) have a notable property
of having no additional constant term (besides $C_{2,3}$) in the expansion
series near $x_3$ and $x_2$ respectively, i.e. we have :
\begin{align}\notag\label{Gj_i}
\Gj_{(2)}(x) & = -\frac{2 }{a^{\frac{3}{2}}\, d_{21}^{\strut \frac{3}{2}} d_{32}^{\frac{3}{2}} \,
\sqrt{\vphantom{|}\xi_2}} \, + \, C_2 +
\frac{3(d_{21} {-} d_{32})}{a^{\frac{3}{2}}\, \strut d_{21}^{\frac{5}{2}}
d_{32}^{\frac{5}{2}}}  \, \sqrt{\xi_2}
 \, + O(\xi_2^{\frac{3}{2}})  \cmma
\\[0.5em]
\Gj_{(3)}(x) & =\hphantom{-} \frac{2 }{a^{\frac{3}{2}}\, d_{32}^{\strut \frac{3}{2}} d_{31}^{\frac{3}{2}} \,
\sqrt{\vphantom{|}\zeta_3}} \, + \, C_3 +
\frac{3(d_{32} {+} d_{31})}{a^{\frac{3}{2}}\, \strut d_{32}^{\frac{5}{2}}
d_{31}^{\frac{5}{2}}}  \, \sqrt{\zeta_3}
 \, + O(\zeta_3^{\frac{3}{2}})  \cmma
\end{align}
where $C_2$ and $C_3$ are exactly the same as in (\ref{gj-intexplicit-C2}), (\ref{gj-intexplicit-C3}).

Sometimes it is also convenient to choose the integration constant so we have $\Gj(0)=0$ :
\begin{align}
\tilde\Gj(x)=\Gj(x)-\Gj(0) \cmmb
\end{align}
In general, the choice of this constant is tantamount to the supertranslation by $\alpha=\frac{C\sin\Theta}{A\K}$ (see
\ref{uhat}), i.e. to the corresponding redefinition of $\hat u$. Note that this is still compatible with our choice of $\hat u$,
namely preserving $\hat u =0$ on
$\scri^+$ where the black hole reaches it (see figure \ref{slices}).

\section{Bondi mass change positivity in $\bold u \rightarrow \pm 0$ limit}
\label{A-bondimass}

In order to prove positivity of the first term in the series in $u$, around $u=0$, of the
Bondi mass $M(u)$ for all $\alpha \sim \sin\Theta$ supertranslations,
we have to prove that the following quantity
\begin{align}
 m_{2} \equiv -\frac{(x_1{+}x_{3}{-}x_{2})\,\K_{2}}{2A^2 (x_{3}{-}x_{2})(x_{2}{-}x_1)} -
 \bigg [ \frac{24{-}3\,\K_{2}^2 G'^2}{96A^2G}+\frac{G'''x\K_{2}^2}{24A^2} - \frac{C_{2}^2}{8A^2\K_{2}^2} \bigg ]_{|x=x_0} \cmma
\end{align}
is positive\footnote{Being zero at one of the boundary points and, as can also be shown, diverging at the other.} for all $x_0$ in the open interval $I \equiv (x_2,x_3)$. First, we prove that the function
\begin{align}\notag
h= \frac{1}{4A^2\sqrt{G}} \ \Big[\ \frac{1}{4}\K_2^2 GG'G'' - \frac{1}{6} \K_2^2 G^2G''' - \frac{1}{8} \K_2^2 G'^3 + G' \ \Big] -
\frac{C_2}{4A^2\K_2^2}
\end{align}
is positive for all $x_0 \in I$. Realizing that $C_2=-\tilde\Gj_{(2)}(x_0)$ where $\tilde\Gj_{(2)}$ is $\Gj_{(2)}$ with $C_2{=}0$
(\ref{gj-intexplicit-C2}), we see that together with (\ref{CMcon}) and (\ref{Gj_i}) this gives us the
limiting value $h_{|x=x_0=x_2} \equiv h(x_2)=0$.
Now, to prove that $h> 0$ it
is sufficient\footnote{In fact, it still suffices if $h_{,x}$ is zero at finite number of points, as in our case,
where this happens at one of the boundary points.}  to show that $\ds \frac{d h}{d x}_{|x=x_0} > 0$. But this is obvious, since :
\begin{align}\notag
\frac{d h}{d x}_{|x=x_0} = \frac{1}{A^2\K_2^2G^{\frac{3}{2}}} \ \Big( \ \frac{1}{2}+\frac{1}{4}\K_2^2GG''-\frac{1}{8}K_2^2G'^2 \, \Big)^2
\cmmb
\end{align}
To conclude the proof, we realize that $\ds h=G^{\frac{3}{2}}\,  \frac{d m_2}{d x}$. Since $m_2(x_2)=0$ (see \ref{Mu0_limit}),
this means that $m_2>0$ for all $x_0 \in (x_2,x_3)$. The proof of the second case, for the regular axis segment $x=x_3$ is completely analogous;
the function $h$ is the same, only now $h(x_3)=0$. Hence $\ds \frac{d h}{d x}_{|x=x_0}>0$ implies $h<0$ on $I$, which together with
$m_3(x_3)=0$ again implies $m_3>0$ on $I$.

\vskip 2cm


\nopagebreak

\end{document}